\documentclass[a4paper,11pt]{article}
\usepackage[dvips]{graphicx}
\newcommand{\rf}[1]{(\ref{#1})}
\newcommand{\beq}{\begin{equation}}
\newcommand{\eeq}{\end{equation}}
\newcommand{\beqr}{\begin{eqnarray}}
\newcommand{\eeqr}{\end{eqnarray}}
\newcommand{\ba}{\begin{array}}
\newcommand{\ea}{\end{array}}
\newcommand{\bitem}{\begin{itemize}}
\newcommand{\eitem}{\end{itemize}}
\newcommand{\p}{\partial}
\renewcommand{\inf}{\infty}
\renewcommand{\Re}{\textrm{Re}}
\renewcommand{\Im}{\textrm{Im}}
\renewcommand{\a}{\alpha}
\renewcommand{\b}{\beta}

\newcommand{\G}{\Gamma}
\renewcommand{\d}{\delta}
\newcommand{\ep}{\epsilon}
\newcommand{\ve}{\varepsilon}
\renewcommand{\th}{\theta}

\newcommand{\Sc}{Schr\"odinger\,}
\renewcommand{\l}{\lambda}
\renewcommand{\L}{\Lambda}
\newcommand{\x}{{\bf x}}
\newcommand{\y}{{\bf y}}
\newcommand{\z}{{\bf z}}
\newcommand{\q}{{\bf q}}
\newcommand{\pp}{{\bf p}}
\newcommand{\kk}{{\bf k}}
\newcommand{\hq}{\hat{\bf q}}
\newcommand{\hp}{\hat{\bf p}}
\newcommand{\hH}{\hat{H}}

\newcommand{\f}{\varphi}
\newcommand{\tf}{\tilde{\varphi}}
\newcommand{\hf}{\hat{\varphi}}
\newcommand{\df}{\dot{\varphi}}
\newcommand{\om}{\omega}

\newcommand{\da}{\dagger}
\newcommand{\dis}{\displaystyle}
\newcommand{\um}{\frac{1}{2}}
\newcommand{\lag}{{\cal L}}
\newcommand{\D}{{\cal D}}
\renewcommand{\P}{\hat{\cal P}}

\newcommand{\ra}{\rightarrow}
\newcommand{\nn}{\nonumber}

\def\Journal#1#2#3#4{{#1} {\bf #2} (#3) #4}

\def\NPB{{\em Nucl. Phys.} B}
\def\PLB{{\em Phys. Lett.}  B}

\def\PRD{{\em Phys. Rev.} D}

\begin{document}
\baselineskip=15pt
\begin{titlepage}
\vspace*{-1.2cm}
\flushright{FTUV/98-39\\IFIC/98-40}
\flushright{DTP 98/57}
\vspace{0.7cm}
\begin{center}
{\Large{\bf Finite VEVs from a Large Distance Vacuum Wave Functional
}}\\

\vspace{1.7cm}

{\large Alfonso\ Jaramillo $^{a}$}\\

\vspace{0.2cm}
{Departamento de F\'{\i}sica Teorica and I.F.I.C.}\\
{Centro Mixto Universidad de Valencia -- C.S.I.C.}\\
{E-46100 Burjassot (Valencia), Spain.}

\vspace{0.9cm}
{\large Paul\ Mansfield $^b$}\\

\vspace{0.2cm}
{Department of Mathematical Sciences}\\
{University of Durham}\\
{South Road}\\
{Durham, DH1 3LE, England.}

\vspace{2.0cm}
{\bf Abstract}
\vspace{0.2cm}
\begin{quotation}
{\small \noindent
We show how to compute vacuum
expectation values from derivative expansions of the
vacuum wave functional. Such expansions appear to be valid
only for slowly varying fields, but by
exploiting analyticity in a complex scale parameter
we can reconstruct the contribution from
rapidly varying fields.}
\end{quotation}

\end{center}

\vspace{2.0cm}
\flushleft{
$^a$ Alfonso.Jaramillo@uv.es\\
$^b$ P.R.W.Mansfield@durham.ac.uk\\
}
\end{titlepage}


\section{Introduction}
Although canonical quantisation provides the basic formalism of
quantum field theory, the corresponding Schr\"odinger Representation,
in which the field operators are diagonal, has not received
commensurate attention. This is partly due to the popularity of the
functional integral which displays space-time symmetries manifestly,
and partly because the existence of wave functionals was only shown by
Symanzik as late as 1981, \cite{symanzik}. Nonetheless there has been
growing interest in the subject as a result of the search for new tools
in field theory, and also because the \Sc Representation is implicit in much recent work on field theories defined on space-times with boundaries,
see for example \cite{witten}.

The vacuum wave-functional (VWF), $\Psi_0$, may be constructed as a functional
integral over the Euclidean space-time $x^0<0$ which has the
quantisation surface $x^0=0$ as boundary. $W\equiv \log (\Psi)$ is then
a functional of the values the field takes on the  boundary. As we will
see, these boundary values act as source terms in the functional
integral. Symanzik showed that, at least in perturbation theory, this
functional has a finite limit as the cut-off is removed, subject to the
inclusion of the usual counter-terms together with additional ones
localised to $x^0=0$ which result in the boundary values of the field
undergoing an additional field renormalisation, \cite{symanzik}.  These boundary
counter-terms are absent in fewer than three dimensions, as they are for
Yang-Mills theory in four dimensions due to gauge invariance. He also
proved the existence of the Schr\"odinger equation for $\phi^4$ in four
dimensions.

In \cite{Paul} it was shown that the vacuum functional of the
scaled Yang-Mills field ${\bf A}^s({\bf x})\equiv {\bf A}({\bf x}/
\sqrt s )/\sqrt s$ extends to an analytic function of $s$ in the complex
$s$-plane with the negative real axis removed. This also applies to
scalar field theory. This allows the vacuum functional to be
reconstructed for arbitrary ${\bf A}({\bf x})$ in terms of the scaled
field ${\bf A}^s({\bf x})$ for large $s$ using Cauchy's theorem. The
scaled field is slowly varying for large $s$, and for such a field we
would expect to be able to expand $W$ in powers of derivatives divided
by the lightest glueball mass (in our appendix C we give an
argument to justify the possibility of performing this local
expansion). Thus $W$ can be obtained for arbitrary
$\bf{A}({\bf x})$ from a knowledge of this derivative expansion. The
existence of this expansion was originally considered by Greensite
\cite{halp} who found the leading terms from a Monte-Carlo simulation
of lattice gauge theory.  We emphasize that this procedure is valid for
arbitrary ${\bf A}({\bf x})$, it does not amount to restricting
attention to slowly varying field configurations, it is
simply a way of parametrising the function space of sources
in the functional integral in terms of a derivative expansion.

$\Psi$ satisfies the \Sc equation, but this takes a special form
when using the derivative expansion for $W$ due to the
employment of Cauchy's theorem. This was described
in \cite{Paul} where a non-perturbative
approximation scheme was outlined.
The semiclassical expansion of this
equation was shown to agree with the direct semi-classical
evaluation of $W$ via Feynman diagrams in \cite{Marcos}, and
extended to Yang-Mills theory in \cite{Marcos2} where the
beta-function was correctly reproduced from the derivative expansion.
This is not as obvious as it might appear because a naive
insertion of a local expansion into the usual \Sc equation will
not converge for momenta greater than the mass of the lightest
particle and so will not lead to the correct behavior
as the ultra-violet cut-off in that equation is removed.
What was missing from this work was a method for constructing vacuum
expectation values (VEVs) directly from the derivative expansion,
which is the subject of this paper. We will show that when
these are written as functional integrals over the  boundary values of
fields they are analytic in an ultra-violet momentum cut-off in the
plane cut as above. Again, Cauchy's theorem may be used to compute VEVs
for large cut-off from a knowledge of the corresponding functional
integral for small cut-off, which in turn can be obtained from the
derivative expansion, or some other systematic approach. Notice that if we try to compute the VEV in the most obvious way, by expanding the logarithm of the vacuum functional
in a local expansion and doing the usual perturbative approach then we
would get a sum of contractions which would, in general, lead to
ultra-violet divergences of all orders.  These divergences cannot be
absorbed into renormalisation of the wave-functional to obtain finite VEVs because the
wave-functional is already finite according to Symanzik's work.
(The only possibility for such renormalisation is if the inner product involves
a non-trivial weight functional with coefficients that can be chosen to cancel
divergences. The form of these weights is determined by the
hermiticity of the Hamiltonian operator, and is very restricted.
In 1+1 dimensional scalar field theory, and Yang-Mills theory such
weights are absent.)
The origin of
these ultra-violet divergences is that we would be attempting to compute the integral for field
configurations beyond the convergence radius of the derivative
expansion, and this is inconsistent.
In this paper we propose a method for computing VEVs
in which the same expansion is employed but with a cut-off
that lies inside the convergence radius of the series.
Typically this means that the cut-off is smaller than the mass of the
lightest particle. It therefore does not appear to be an ultraviolet cut-off.
However we will be able to send the cut-off to infinity
in this expansion because, as we will show, that VEVs
are analytic in the cut-off when we continue to complex values. Thus
we can use Cauchy's theorem to relate the large cut-off behaviour which we need to compute, to the small cut-off behaviour which we can calculate using the
local expansion of the funcctional integral.

We concentrate throughout on the toy model of $\phi^4$-theory
in 1+1 dimensions as this is particularly straightforward given the
absence of boundary counter-terms resulting in there being
no wave-function renormalization. The absence of boundary
counter-terms is shared by Yang-Mills theory which also has
a correspondingly simple \Sc equation. Super-renormalizable
theories are, in any case, of interest in their own right by virtue of
their connection with integrable theories and with String Theory.
We will only discuss the VEV of operators which will be diagonal in
field configuration space (we do not expect that our
conclusions will change if we consider more general operators,
$A(\pi,\phi)$, as we can see in \cite{Paul} where the
analyticity of $H\Psi_0$ is shown). Finally, with an analytical continuation it is often
difficult to estimate truncation errors, but we will see that they can be
controlled.

Several authors \cite{symanzik,cornwall,suranyi,osborn,LNWW,Fradkin}
have devised perturbative and non-perturbative
aproaches to compute the vacuum wave-functional.
In section 2 we describe its representation
as a functional integral. In the next section we give a general
discussion of the construction of VEVs in the \Sc Representation in
terms of Feynman diagrams. We display the mechanism whereby the Feynman
diagram expansion of $W$, which makes use of a propagator on the
space-time $x^0<0$ with Dirichlet boundary conditions, leads to the
usual  Feynman diagrams for VEVs on the full space-time with the
standard propagator. We end the section by giving an operator approach which
uses the results of appendix A.  In section 4 we translate the calculations of the previous section into the language of first quantisation in which the vacuum functional
can be expressed in terms of random paths that are reflected
at the quantisation surface. In section 5 we describe the analyticity of VWF and
VEVs and describe the resummation of the series in the cut-off.
Section 6 ilustrates our method in the simpler context of
non-relativistic quantum mechanics and we have left to the appendix B
mathematical details of our method of analytic continuation.
In the section 7 we will discuss
the computation of the equal-time two point function through diagrams in a
dimensionally reduced effective theory.
The last section is devoted to our conclusions.


\section{Representations for the vacuum wave functional}
The VWF is the inner product $\langle\,\f\,|\,0\,\rangle$ of
the vacuum $|\,0\,\rangle$ and an eigenbra of the field operator $\hat{\phi}$
restricted to the quantization surface (which we take to be $t=0$)
belonging to the eigenvalue $\phi(\x)$:
\beq
\langle\,\f\,|\hat{\phi}(\x,0)=\f(\x)\langle\,\f\,|
\eeq
The $\f$-dependence of the eigenbra may be made explicit by
writing
\beq
\langle\,\f\,|=\langle\,D\,|\exp(i\int d\x\,\,\f(\x)\hat{\pi}(\x,0))
\label{eq2}
\eeq
where $\langle\,D\,|$ is annihilated by $\hat{\phi}(\x,0)$, i.e. it is
the state $\langle\,\f\,=0|$, D stands for Dirichlet, and $\hat{\pi}$ is
canonically conjugate to  $\hat{\phi}$. The canonical commutation
relations then yield \rf{eq2} together with
\beq
\frac{\d}{\d\f(\x)}\langle\,\f\,|=i\langle\,\f\,| \hat{\pi}(\x)
\eeq
if we apply the Euclidean time evolution operator $\exp(-\hat{H}T)$ to
any state, $|\,\upsilon\,\rangle$, not orthogonal to the vacuum, then for large times
\beq
\exp(-\hat{H}T)|\,\upsilon\,\rangle\sim |\,0\,\rangle e^{-E_0T}\langle\,0\,|\,\upsilon\,\rangle\,\,\,\,(T\rightarrow \inf)
\eeq
where $E_0$ is the energy of the vacuum. Thus
\beq
\Psi[\f]= \lim_{T\rightarrow \inf} N \langle\,D\,|e^{i\int d\x\,\, \f(\x)\hat{\pi}
(\x,0)}e^{-\hat{H}T}|\,\upsilon\,\rangle
\label{eq5}
\eeq
Where $N$ is a normalization constant depending on $|\,\upsilon\,\rangle$. Using
$\hat{\pi}=\frac{\p\hat{\f}}{\p t}$ this, as we will explicitly show later,
 may be written as the functional
integral
\beq
\int\D\phi\,\, e^{-S_E[\phi]-\int d\x\,\,(\f(\x)\dot{\phi}(\x,0)+
\L \f^2(\x))}
\label{eq6}
\eeq
where $S_E$ is the Euclidean action for the space $t\le 0$. $\L$ is
a regularization of $\d(0)$ that arises from the differentiation of
the time ordered product that is represented by the functional
integral, i.e.
\beq
T(\hat{\pi}(\x,t)\hat{\pi}(\x',t'))=\frac{\p^2}{\p
t\p t'}T(\hat{\f}(\x,t) \hat{\f}(\x',t'))-i\d(\x-\x')\d(t-t')
\label{eq7}
\eeq
On the boundary $t=0$ the integration variable $\phi$ should vanish,
reflecting the fact that $\langle\,D\,|\hat{\phi}(\x,0)=0$.

Alternatively, we can obtain this path integral
representation for the VWF, by beginning with
\beq
\Psi_0[\phi]=N\lim_{\tau\rightarrow \inf}e^{\tau E_0}
\int\D\f(\x,t)\,\, e^{-S_E(\f,\df)}{\Big |}_{\f(\x,0)=\phi(\x)}
\label{FeyVWF}
\eeq
where $\f(\x,-\infty)$ can be anything,
and performing a functional change of variables
in the path integral in such a way that we do not have
the $\phi(\x)$-field dependence in the integration limit.
Our change of variables is formally
\beq
\begin{array}{l}
\tilde\f(\x,t) = \f(\x,t)-2\th(t)\phi(\x) \\
\\
{\cal D}\tilde\f = {\cal D}\f \\
\\
{\dis
S_E[\f,\dot{\f}] = \int_{-T}^0 dt\! \int\! d\x \,\,
\lag(\tilde\f+2\th(t)\phi , \dot{\tilde\f}+2\d(t)\phi)}
\end{array}
\label{shift}
\eeq
where $\theta$ is the step function and we take $\th(0)=\um$.  Naively the
$\th$ terms do not contribute to the potential.  Therefore our path integral on removing the tilde
can be written as

\beq
\int {\cal D}\f(\x,t) \,
e^{-S_E(\f,\dot{\f}+2\d(t)\phi)}{\Big |}_{\f(\x,0)=0}
\eeq
In the scalar case we will have
\beq
S_E= \int_{-\inf}^0 dt \int d\x \{\um(\dot{\f})^2 + V(\f)\} +
\int d\x
\,\,\dot{\f}(\x,0)\phi + \int d\x\,\, \d(0)\phi^2
\label{newact}
\eeq
Therefore, the logarithm of the VWF ($W[\phi]$) is given by the sum
of connected diagrams constructed from the new action \rf{newact} and
with a boundary at $t=0$ where the field vanishes.
The new
action contains a source term on the boundary and a $\delta(0)$
term (to be regulated) coupled to the the boundary fields.
This argument has been rather too formal. To be more careful we should
smooth the $\th$
functions in \rf{shift}, taking them to be non-constant in a region of size $1/\L$,
and this will regulate $\delta (0)$. So we replace $\th$ by
$\th_\L$.
With a cutoff this function will be given by
\beq
\th_\L (t) = \frac{i}{2\pi}\int_{-\L}^{\L}d\om \frac{1}
{\om+i\ep}e^{-i\om t},
\eeq
and we have $\th_\L(t)= \th(t)-\frac{1}{\pi}
\frac{\cos(\L t)}{\L t}(1+O(\frac{1}{\L t}))$.
If we had already a cutoff in space-time
then the $\th$ functions will be regulated by this cutoff. Therefore, because
of the appearance of the $\d(0)$ terms, it is necessary to regulate both
the space-like and the time-like dimensions. If we
regulate the space-like dimensions keeping unregulated the time-like
direction (for instance using dimensional regularization for the
space-like dimensions, as is done by Symanzik \cite{symanzik}) then
we need to introduce another regulator for the time direction. Assuming
that we have a cutoff in space-time, then  the terms proportional to
some power of  $\th_\L$,
in $V(\f+2\th_\L(t)\phi)$, will  vanish when
$\L\rightarrow\inf$ because the time integral (which occurs in the definition
of the action) will only be non-zero in a region of size $1/\L$ around
the endpoint $t=0$. When we compute the perturbative quantum corrections
we may get some divergent loop diagram that may compensate for the vanishing
contribution of the $\th_\L$-terms insertions. But for that to happen we
need a linear
divergence (because $\th_\L(t)\sim 1/\L$) or a time derivative
acting on the $\th_\L$. In $1+1$ dimensions we do not get linear
divergences and in our case (where we do not have derivative
interactions) we do not have vertices with both  $\th_\L$ and $\df$,
so we can ignore such terms.

We can also see how to get
\rf{eq6} from \rf{eq5} together with the $\th_\L$-terms within the
canonical operator formalism.
We will use the following identity
\beq
e^Ae^B=Te^{\int_{-1}^0 dt\,\, e^{Bt}(B+A)e^{-Bt}}
\label{prod-exp}
\eeq
where $A$ and $B$ are arbitrary matrices (or operators) and $T$ is the
time-ordering operator. As usual, the $T$-ordering implies that the first term in the exponential is to
be thought of as $B(t)$ and then,
at the end set to a constant. This relation can be derived by considering the operator $U(t)\equiv
e^{At}e^{Bt}$, then we calculate $\frac{d}{dt}U(t)$
and then we integrate it back to get
the integral equation $$ U(t)=1+\int_0^t dt'\,\, U(t')\,(B+A(-t')) $$
with $A(t)\equiv e^{Bt}Ae^{-Bt}$. Once we solve the integral equation
in terms of a $T$-ordered exponential, we set $t=1$ and we get
\rf{prod-exp}.

Now we will consider
\beq
\langle\,D\,|e^{i\int d\x\,\, \f(\x)\hat{\pi}
(\x,0)}e^{-T\hH}|\,\upsilon\,\rangle=\langle\,D\,|e^{i\int d\x\,\, \f(\x)\hat{\pi}
(\x,0)}e^{-\ep\hH}e^{-(T-\ep)\hH}|\,\upsilon\,\rangle
\label{split}
\eeq
and we will use our relation  \rf{prod-exp}
to combine the $e^{i\int d\x\,\, \f(\x)\hat{\pi}
(\x,0)}e^{-\ep\hH}$ term into a single exponential, but before that
we will follow some intermediate steps. Firstly we take (and to shorten
the notation $\int d\x\,\, \phi(\x)\hat{\pi}
(\x,0)\equiv \phi \hat{\pi}$)
\beq
e^{-\ep\hH}e^{i\phi\hat{\pi}}=Te^{-\ep \int_{-1}^0 dt\,\,
(\frac{-i}{\ep}\phi\hat{\pi} + e^{i\phi\hat{\pi}t}\hH
e^{-i\phi\hat{\pi}t})}=Te^{
-\int_{-\ep}^0 dt\,\, (\frac{-i}{\ep}\phi\hat{\pi} +
\hH(\hat{\pi},\hat{\f}+\phi\frac{t}{\ep}))}
\eeq
which (after hermitian conjugate, $\phi\rightarrow -\phi$, $t\rightarrow
-t$ and later $t\rightarrow t-\ep$) will become
\beq
e^{i\phi\hat{\pi}}e^{-\ep\hH}=Te^{
-\int_{-\ep}^0 dt\,\, (\frac{-i}{\ep}\phi\hat{\pi} +
\hH(\hat{\pi},\hat{\f}+(\frac{t}{\ep}+1)\phi))}
\eeq
Therefore
\beq
\langle\,\phi\,|e^{-T\hH}|\,\chi\,\rangle=\langle\,D\,|e^{i\phi\hat{\pi}}e^{-T\hH}|\,\chi\,\rangle=
\langle\,D\,|Te^{-\int_{-T}^0dt'\,\,\hH'(t')}|\,\chi\,\rangle
\eeq
with $\hH'(t)\equiv \hH(\hat{\pi},\hat{\f}+2\th_\ep(t)\phi)
-2i\d_\ep(t)\phi\hat{\pi}$, where
the $\th_\ep(t)$ is defined by
$$\th_\ep(t)=\left\{
\ba{ll}
0& \textrm{if $t\le-\ep$}\\
\frac{t}{2\ep}+\um & \textrm{if $-\ep < t < \ep$} \\
1&\textrm{if $t\ge \ep$}
\ea
\right. $$
and the $\d_\ep(t)$ is obtained by taking its derivative. Notice that
we needed to put a factor $2$ in front because $\int_{-\ep}^0 dt\,\,
\frac{1}{\ep} =1=2\int_{-\ep}^0 dt\,\, \d_\ep(t)$.
With such a definition of the Hamiltonian
(we have $\hH(t)=\hH$ for $t<-\ep$) the time evolution until $t<-\ep$
is reproduced by the third exponential in the r.h.s. of eq.
\rf{split}).  As we see, we have the same result
that we obtained by shifting the integration variable in the path-integral.
The path-integral representation now follows from the standard
construction. Just the last step ($t<-\ep$) is
different from usual. Let us illustrate it for the case of quantum mechanics
\beqr
\langle\,q\,|e^{-\ep(\um
\hp^2+V(\hq))}|\,q_1\,\rangle & = & \langle\,q=0\,|e^{-\ep(\um \hp^2 -\frac{i}{\ep}q\hp
+V(\hq+2\th_\ep(t)q))}|\,q_1\,\rangle=\nn\\
 &=&N\, e^{-\um
(\frac{-q_1}{\ep}+\frac{q}{\ep})^2-V(q_1+2\th_\ep(t)q)}=N \,
e^{-S(q,\dot q)}
\eeqr
where the $\frac{-q_1}{\ep}$ (in the last  equality) is interpreted as
$\dot{q}_1$ and $\frac{q}{\ep}$ as $2\d_\ep(t) q$. As we see we get the
same action as before.  Finally we can write the VWF as
\beqr
 \Psi[\f] & = & \lim_{T\rightarrow \inf} N \langle\,D\,|Te^{-\int_{-T}^0dt'\,\,
\hH(t')}|\,\phi\,\rangle=\nn\\
 &= & \lim_{T\rightarrow \inf} N
\int_{\phi,t=-T}^{0,t=0} \D \chi(\x,t)\,\,
e^{-S_E(\chi(\x,t)+2\th_\ep(t)\f(\x), \, \dot{\chi}(\x,t)
+2\d_\ep(t)\f(\x))}
\eeqr
Notice that we could have got the following relation
\beq
e^{i\phi\hat{\pi}}e^{-T\hH}=Te^{
-\int_{-T}^0 dt\,\, (\frac{-i}{T}\phi\hat{\pi} +
\hH(\hat{\pi},\hat{\f}+(\frac{t}{T}+1)\phi))}
\eeq
which could be derived with \rf{shift} using $2\th(t)=1+t/T$. Lastly, we may
also construct a relation
\beq
e^{i\phi\hat{\pi}}e^{-T\hH}=Te^{
-\int_{-T}^0 dt\,\, (\hH-\frac{i}{T}\phi\hat{\pi}(t))}
\eeq
with $\hat{\pi}(t)=e^{-t\hH}\hat{\pi}e^{t\hH}$ but this time we cannot use
\rf{shift} to obtain the corresponding path integral version.
  We have seen from a range of methods that the (time-independent) VWF
will be given by
\beq
\Psi[\phi(\x)]=N \int\D \f(\x,t)\,\, e^{-S_E-
\int d\x\,\dot\f(\x,0)\phi(\x)-\int d\x\,\d_\ep(0)\phi^2}{\Big |}_{\varphi(\x,0)=0}
\label{SymVWF}
\eeq
We can also obtain a path integral representation for the VWF
(with $\phi$-independent boundary conditions) without any delta
function by using a different shift in the field in \rf{shift}.
Consider shifting $\f$ by a solution to the free field equations
denoted by $\Omega_\L(t)\phi$, where $\Omega_\L(t)$ will be defined by
\beq
\Omega_\L(t)=\frac{\sinh (t\sqrt{-\p^2+m^2})}{\sinh (\L\sqrt{-\p^2+m^2})}
\label{th-cornwall}.
\eeq
Thus $S_0(\f+\Omega_\L(t)\phi)= S_0(\f)+S_0(\Omega_\L(t)\phi)$
which yields

\beqr
&&
\int\D \f\,\, e^{-S_0(\f)-\int V(\f)}{\Big |}_{\varphi(\x,0)=0,\varphi(\x,\Lambda)=\phi (\x)}\nn\\
&&
=
\int\D \f\,\, e^{-S_0(\f)-\int V(\f+\Omega_\L(t)\phi)
-S_0(\Omega_\L(t)\phi)}{\Big |}_{\varphi(\x,0)=\varphi(\x,\Lambda)=0}.
\label{cornwall}
\eeqr
This provides an alternative derivation of the starting point used in \cite{cornwall}
to set up a non-perturbative algorithm to
compute the VWF by resumming the perturbative expansion of
\rf{cornwall}.
 Another possibility is to shift the field by a solution to the
Euler-Lagrange equations of the full action
with boundary
conditions $\phi'(x,0)=\phi$ and $\phi'(x,-T)=0$ (where
$T\rightarrow \inf$). In the background field technique a
$\phi'(x,t)=\langle\hf(x,t)\rangle_{\phi}$ (the expectation value is taken with the
boundary condition $\f(x,0)=\phi$) is taken (which at first order
coincides with the one which minimises the action) and then the $\Psi_0$
gives the exponential of the effective action evaluated at the
configuration $\phi'$. This is the method used in \cite{LNWW}  where
the gauge theory case is analyzed and $\phi'$ plays the role of their
induced background field, although they do not expand in the background field,
and in \cite{osborn} where they study general
scalar theories with curved boundaries using dimensional regularization
for the space-time dimensions. In this way they give a method to use
dimensional (space-time) regularization with the \Sc representation
where the time is given by the coordinate perpendicular to the
boundary of the $d$-dimensional space-time. Note that usually
\cite{symanzik} the space-like dimensions are regulated differently
(for example with dimensional regularization) from the time (for
example by time-splitting).

We end this section by deriving the important path integral
representation introduced by Symanzik \cite{symanzik}, where the Dirichlet
boundary
conditions are only introduced through a boundary term. Again we can
derive this representation in an easy way, following steps similar to
those of eq. \rf{shift}. Consider (as before, we understand the delta
functions to be regularized)
\beq
\int\D\f(\x,t)\,\, e^{-\int\limits_{-\inf}^0 dt \int
d\x \{\um(\dot{\f})^2 + V(\f)\}
}\,\,\frac{\dis \int\D C(\x)\,\,
e^{\int d\x  \d(0)(\phi-C)^2}}
{\dis \int\D C(\x)\,\, e^{\int d\x
\d(0)(\phi-C)^2}}\eeq
where $\varphi(\x,0)=\phi (\x)$ and $\varphi(\x,-\infty)=0$.
Now absorb the denominator into a proportionallity constant and
commute the $\D\f$ and $\D C$ integrals. Shifting
the variables $\tilde\f(\x,t) = \f(\x,t)-2\th(t)\phi(\x)+2\th(t)C(\x)$
gives
\beq
N\!\!\int\D C(\x)\,D\f(\x,t)\,\, e^{-\!\!\int\limits_{-\inf}^0 \!\!\! dt\!\! \int
\!\! d\x \{\um
\dot{\f}^2+V(\f)+2\dot{\f}\d(t)(\phi-C) +2 (\d(t))^2\, (\phi-C)^2\} }
e^{\int \!\! d\x\, \d(0)(\phi-C)^2}
\eeq
where now $\varphi(\x,0)=C (\x)$ and $\varphi(\x,-\infty)=0$.
After the cancellation of the exponential our VWF will be given by \beq
\Psi_0[\phi]=N'\,\int\D C(\x)\,\D\f(\x,t)\,\, e^{-\int\limits_{-\inf}^0 dt \int d\x \{\um\dot{\f}^2+
V(\f)\}-\int d\x \,\dot{\f}(\x,0) (\phi(\x)-\f(\x,0))}\,,
\eeq
where the $\int d\x$ integral has to
be understood as $\int dt\, d\x 2\d(t)$.
We can now interpret $\int\D C(\x)\,\int\D\f(\x,t)$
with the conditions $\varphi(\x,0)=C (\x)$ and $\varphi(\x,-\infty)=0$ as an ordinary $\int \D\f(\x,t)$ with $t\le 0$ and free
boundary at $t=0$. This formula agrees with Symanzik's one (eq. (2.10) of
\cite{symanzik}) but for an arbitrary potential. We expect that quantum
corrections will renormalize it.

As Symanzik has discussed, placing source terms on the boundary leads to divergences, \cite{symanzik}.
These appear in perturbation theory because the field is placed at the same point as the image charges that enforce the boundary conditions on propagators.
In order to regulate these divergences we should
split in time the fields (thus the fields in
\rf{corr} will be defined at different ordered times, but with their
$|t|<\ep$). These divergences appear as the coefficient of local
operators forming a {\it boundary operator expansion}
\cite{osborn,diehl}
(analogous to the
usual operator product expansion). These coefficients will scale with
some non-trivial dimension (in perturbation theory they are calculated
as a series in the coupling), which (order by order in the coupling)
will lead to logarithms of $t$. In a perturbatively renormalizable
theory (where the anomalous dimensions cannot make relevant an
irrelevant operator) we need to only consider the operators of
dimension smaller or equal than the product of operators at the
boundary because $\ep\rightarrow 0$. Therefore,
in order to use the \Sc representation (where $\ep=0$ is implied), we
should subtract the previous divergences (Wilson-boundary
coefficients) in the original lagrangian.  Extension of the
validity of the boundary operator expansion to the
non-perturbative domain suggests that we may have to consider
additional relevant fields, we assume that this is not the case.
Symanzik \cite{symanzik} showed that in $\phi^4$ theory in $3+1$
dimensions only two counterterms where needed ($\phi^2$ and
$\phi\p_0\phi$) and conjectured that in a general perturbatively
renormalizable field theory in four dimensions we only need
operators of dimension less or equal than three. Because in lower
dimensions we find fewer UV infinities, we will limit our discussion to
scalar $\phi^4$ theory in $1+1$ dimensions, where there is no
wave-function renormalization and \rf{corr} is valid.


\section{Feynman diagram expansion of VEVs}
The purpose of this section is to describe how the Feynman
 diagram expansion of the VWF, $\Psi_0[\phi]$, generates the usual
diagrams of equal time Green's functions via the following relation
\beq
\langle\,\phi(\x_1,0)\cdot\cdot\cdot\phi(\x_n,0)\,\rangle =
\int \D\phi(\x)\,\,
\phi(\x_1)\cdot\cdot\cdot \phi(\x_n)\,\, |\Psi_0[\phi]|^2
\label{corr}
\eeq
When we use representation \rf{FeyVWF} for $\langle\,\phi|\,\Psi_0\rangle$ in \rf{corr}
we formally obtain the usual path integral for the time ordered product
of field operators, as we should. If instead we first perturbatively compute the vacuum
functionals using \rf{SymVWF} then we
get some unusual diagrams that generate an effective action which will
be used to compute \rf{corr} with new propagators and
(non-local) vertices. It is of interest to see how these combine
to produce the usual result for the VEV. This will also allow us to use ordinary Feynman diagrams to compute
\rf{corr} (which will be used in the section 7 to compute an equal-time
propagator).

Given the comments at the end of the last section we take the Euclidean action
in 1+1 dimensions as
\beq
S_E=\int d\x dt\,\,\bigg(\um(\dot{\phi}^2+\phi'^2+M^2(\L)\phi^2)+
\frac{g}{4!}\phi^4\bigg)
\eeq
In perturbation theory the only divergent diagrams with
external legs are tadpoles which can be removed by normal ordering. This
enables the dependence of $M$ on the cut-off $\L$ to be
calculated. We will regulate with a cut-off on the spatial component of
the  momentum (in $1+1$ dimensions this is almost sufficent, as we will see), so that
\beq
\begin{array}{ccl}
M^2(\L) & = & {\dis M^2+\hbar\d M^2-\frac{g\hbar}
{4}\int_{p^2<\L}
\frac{dp^0}{2\pi}d{\bf p}\,\,
\frac{1}{p^{02}+{\bf p}^2+M^2} =}\\
\\
 & = & {\dis M^2+\hbar\d M^2-\frac{g\hbar}{4}\int_{p^2<\L}
 \frac{d{\bf p}}{
2\pi}\,\,\frac{1}{\sqrt{{\bf p}^2+M^2}}=}\\
\\
 & = & {\dis M^2+\hbar M_c^2}
\label{eq9}
\end{array}
\eeq
The Feynman diagram expansion of the VWF can now be constructed so that its
logarithm, $W[\f]$, is a sum of connected diagrams in which $\f$ is the
source for $\dot{\phi}$ restricted to the boundary $t=0$. The major
difference from usual Feynman diagrams encountered in free space is that
the propagator satisfies Dirichlet boundary conditions, which means that it
should vanish when either end lies on the boundary. Such a propagator
is given by the method of images as
\beq
G(\x,t,\y,t')=G_0(\x,t,\y,t')-G_I(\x,t,\y,t')
\label{D-prop}
\eeq
where $G_0$ is the free-space propagator and
$$ G_I(\x,t,\y,t')=G_0(\x,-t,\y,t')=G_0(\x,t,\y,-t')$$
\begin{figure}[htp]
\centering
\includegraphics[angle=-90,width=0.75\textwidth]{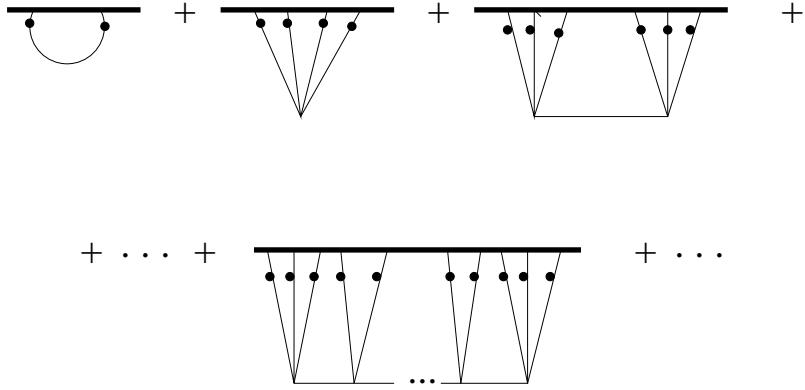}
\caption{Tree level contribution to $\Psi_0[\phi]$.}
\label{fig1}
\end{figure}
The tree-level diagrams that contribute up to the $\f^6$ term in $W[\f]$
are given in figure \rf{fig1}. The heavy line denotes the boundary,
$t=0$, and the dots denote the differentiation with respect to $t$ that
results from $\f$ being coupled to $\dot{\phi}$. When the propagator
ends on the boundary this differentiation leads to $G_0$ and the {\it
image propagator} $G_I$ contributing equally:
\beq
\frac{\p}{\p t}G(\x,t,\y,t')|_{t=0}=2 \frac{\p}{\p t}G_0(\x,t,\y,t')|_{t=0}
\eeq
The $\L\f^2$ term in \rf{eq6} cancels a divergence in the first diagram
of figure \rf{fig1} since this is
\beq
\begin{array}{l}
{\dis \int d\x d\y \,\,\f(\x)\f(\y)\,\,\frac{\p^2}{\p t \p t'}2
G_0(\x,t,\y,t')|_{t=t'=0}}=\\
\\
={\dis \int\frac{d{\bf p}}{
2\pi^2}\,\,\tilde{\f}({\bf p})\tilde{\f}(-{\bf p})\int dp_0 \,\,
\frac{p_0^2}{p_0^2+{\bf p}^2+m^2}}
\end{array}
\eeq
and the $\L\f^2$ term leads to a subtraction so that the
$p_0$ integral is replaced by
\beq
\int dp_0 \,\, \bigg( \frac{p_0^2}{p_0^2+{\bf p}^2+m^2}-1\bigg)=
-\pi \sqrt{{\bf p}^2+m^2}
\label{eq15}
\eeq
All the diagrams that occur in figure \rf{fig1} involve integrals over
the time-like components of Euclidean momenta as this  integration forces
the source terms to be on the boundary, but this is the only
divergent integration since the coupling, $g$, has dimensions of
$M^2$. Another way to deal with this divergence is take the two
times $t$ and $t'$ to be distinct, say $t=\ep<0$ and $t'=0$. Then the last
term of \rf{eq7} is $-i\hbar\d(\x-\x')\d(\ep)=0$ so with this
prescription $\L=0$ and the integral in \rf{eq15} is replaced by
\beq
\int d p_0\,\,\frac{p_0^2}{p_0^2+{\bf p}^2+m^2}e^{ip_0\ep}
\eeq
The exponent allows the $p_0$ contour to be closed in the lower half-plane
giving \rf{eq15}, as before. The $p_i^0$ integrals can be done in a
straightforward way by contour integration (with semi-circle in the upper
plane) using
\beq
\d(\sum_i p_i^0)=\frac{-1}{\pi}\Im \frac{1}{\sum_i p_i^0 +i\ep}
\label{deltaP}
\eeq
Now, we expand $W$ as
\beq
W[\f]=\frac{1}{\hbar}\sum_{n=1}^\inf \int dp_1\cdots dp_{2n}\,\,
\tilde{\f}(p_1)
\cdots \tilde{\f}(p_{2n})\, \G_{2n}(p_1,\cdots, p_{2n})\,
\d(p_1+\cdots+ p_{2n})
\label{eq17}
\eeq
and the tree-level contributions to the kernels, $\G_{2n}'$, are given by
\beq
\begin{array}{l}
{\dis \G_2'(p,-p)=-\frac{1}{4\pi}\sqrt{p^2+M^2}=-\frac{\om(p)}{4\pi}}\\
\\
{\dis \G_4'(p_1,\cdots, p_4)=-\frac{g}{(2\pi)^3 4! (\om(p_1)+
\cdots + \om(p_4))}}
\end{array}
\label{Gtree}
\eeq
and, for $r>2$, the recursion relation
\beq
\begin{array}{l}
\dis\G_{2r}'(p_1,\cdots, p_{2r}) = {4\pi\over\sum_{i=1}^{2r}\om(p_i)}
\sum_{n=2}^{r-1}n(r+1-n)\,\, \times\\
\\
{S\{\G_{2n}'(k,} p_1,\cdots, p_{2n-1})\G_{2(r+1-n)}'(q,p_{2n}
,\cdots, p_{2r})\}
\end{array}
\eeq
where $S$ symmetrises the momenta, $k=-(p_1+\cdots+p_{2n-1})$, and
$q=-(p_{2n}+
\cdots+p_{2r})$.
\begin{figure}[htp]
\centering
\includegraphics[angle=-90,width=0.75\textwidth]{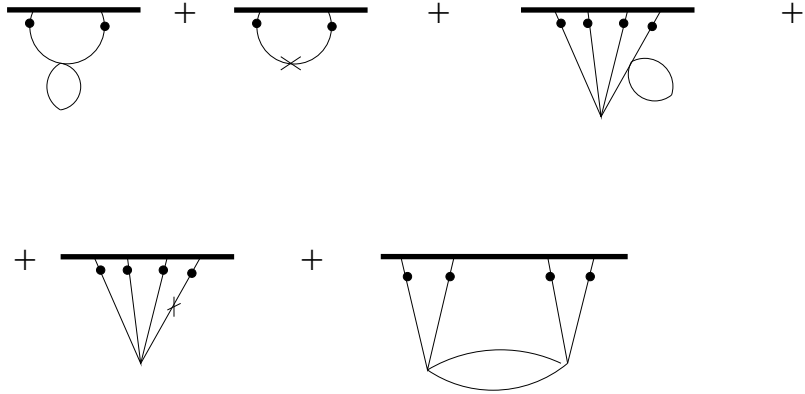}
\caption{One-loop diagrams for $\Psi_0[\phi]$, up to fourth order in
$\phi$.}
\label{fig2}
\end{figure}
The one-loop diagrams up to fourth
order in $\f$ are shown in figure \rf{fig2}, where the cross denotes the
mass counter-term given by the 1-loop term $\d M^2$ in \rf{eq9}, which we
chose to be $\d M^2=\frac{g}{4\pi}$ (so $\G_2^{\hbar}(0,0)=0$). These
yield the $O(\hbar)$ contribution
\beq
\G_2^{\hbar}(p,-p)= \frac{\hbar g}
{32\pi^2p}\sinh^{-1}\bigg(\frac{p}{M}\bigg)-
\frac{\hbar\d M^2}{8\pi\om(p)}
\label{G2}
\eeq
and
\beq
\begin{array}{c}
{\dis \G_4^{\hbar}(p_1,\cdots, p_4)=-\frac{g^2 \hbar}
{(2\pi)^3 4! (\om(p_1)+
\cdots + \om(p_4))}}\\
\\
{\dis S\{ \int_0^\inf \frac{dq}{2\om(q)+
\sum_{i=1}^{4}\om(p_i)}\cdot \bigg( -\frac{1}{\om(q)(\om(q)+
\om(p_1))}}\\
\\
{\dis \frac{3}{(\om(q)+\om(p_1)+\om(p_2)+\om(q+p_1+p_2))}\cdot }\\
\\
{\dis \frac{1}{(\om(q)+\om(p_3)+
\om(p_4)+\om(-q+p_3+p_4))}\bigg)
+\frac{1}{2\om(p_1)\sum_{i=1}^{4}\om(p_i)} \}}
\end{array}
\label{G4}
\eeq
We will now study how the perturbative calculation of $W[\f]$
yields the Feynman diagram expansion of equal time Green's functions
when substituted in \rf{corr}. Keeping only $\G_2'$ in the exponent and
expanding the other contributions to $W[\f]$, which we call
$\tilde{W}[\f]$, yields the Fourier transform of the equal time Green's
functions as
\beq
\begin{array}{c}
{\dis \int d\x_1\cdots d\x_n\,\, \langle\,0\,|\hat{\f}(\x_1,0)\cdots\hat{\f}(\x_n,0)|\,0\,\rangle
\,\, e^{-i\sum_{i=1}^{n}\pp_i\x_i}=}\\
\\
{\dis \int \D\tf \,\, e^{2\int d\pp\,\,\tf(\pp)\G_2(\pp,-\pp)\tf(\pp)}
\sum\frac{(2\tilde{W}[\f])^n}{n!}\, \tf(\pp_1)\cdots\tf(\pp_n)}
\end{array}
\label{eq22}
\eeq
So we have to contract $\tf(\pp_1)\cdots\tf(\pp_n)$ with the $\tf$ in
the diagrams contributing to $\tilde{W}$ using the inverse of $-2\G_2^1$
which is the Fourier transform of the equal time propagator in free-space
(a cut-off in spatial momenta will be implied, but not for the time-like
momenta which will be integrated out)
\beq
\int \frac{d\pp}{2\pi} \,\, e^{i\pp\x}\frac{1}{2\sqrt{\pp^2+M^2}}= \int
\frac{dp_0}{2\pi}\,\frac{d\pp}{2\pi}\,\,\frac{1}
{p_0^2+\pp^2+M^2}e^{i\pp\x}=G_0(\x,0,0,0)
\eeq
\begin{figure}[htp]
\centering
\includegraphics[angle=0,width=0.75\textwidth]{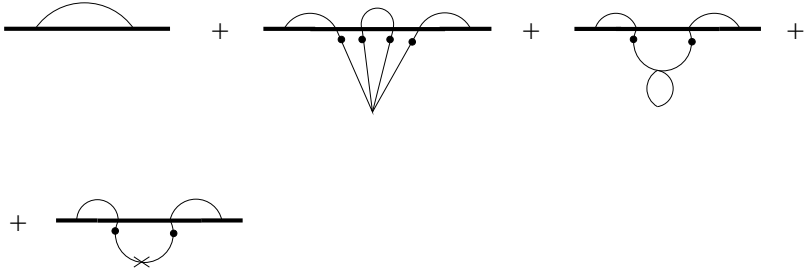}
\caption{Diagrams contributing to the equal time two-point function to
one-loop.}
\label{fig3}
\end{figure}
\begin{figure}[htp]
\centering
\includegraphics[angle=90,width=0.45\textwidth]{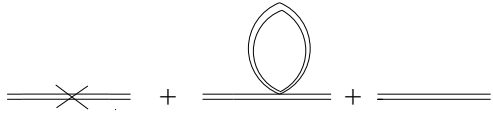}
\caption{Usual Feynman diagram calculation of the two-point function. The
double line is the free space propagator.}
\label{fig4}
\end{figure}
If we denote this by a line above the boundary then the diagrams
contributing to the equal time two-point function are, to one-loop
order, those shown in figure \rf{fig3}, which is to be compared with
the usual Feynman diagram calculation  of the VEV of two
fields in terms of the free space propagator which we denote by a
double line in figure \rf{fig4}. We can understand the equivalence of
these two sets of diagrams, and those of other VEVs, by studying the
{\it gluing together} of two free-space propagators in $D+1$ dimensions on
a $D$-dimensional plane. Consider
\beq
\ba{c}
{\dis \int d\y\,\, G_0(\x_1,t_1,\y,t)\frac{\p}{\p t}G(\y,t,\x_2,t_2)=}\\
\\
{\dis \int \frac{d\pp\, dp_0\,d\q\, dq_0 \, d\y}{(2\pi)^{2(D+1)}}\,\, iq_0\,
\frac{e^{i[\pp(\x_1-\y)+p_0(t_1-t)+\q(\y-\x_2)+q_0(t-t_2)]}}
{(p_0^2+\pp^2+M^2)(q_0^2+\q^2+M^2)}=}\\
\\
{\dis \int \frac{d\pp\, dp_0\,dq_0 \, d\y}{(2\pi)^{(D+2)}}\,\,iq_0\,
\frac{e^{i[\pp(\x-\y)+p_0(t_1-t)+q_0(t-t_2)]}}
{(p_0^2+\om^2(\pp))(q_0^2+\om^2(\pp))}}
\ea
\eeq
The $q_0$ contour may be closed in the upper or lower half-plane,
depending on the sign of $t-t_2$, $\ep(t-t_2)$, giving
\beq
\um \ep(t-t_2)\int\frac{d\pp\, dp_0}{(2\pi)^{D+1}}\,\,\frac{e^{
i[p_0(t_1-t)+\om(\pp)|t-t_2|+\pp(\x-\y)]}}{p_0^2+\om^2(\pp)}
\eeq
and a similar treatment of the $p_0$ integration leads to
\beq
\um \ep(t-t_2)\int\frac{d\pp}{(2\pi)^{D}}\,\,\frac{e^{
i[\om(\pp)(|t_1-t|+|t-t_2|)+\pp(\x-\y)]}}{2\om(\pp)}
\eeq
Given that
\beq
G(\x,t_1,\y,t_2)=\int\frac{d\pp}{(2\pi)^{D}}\,\,\frac{1}{2\om(\pp)}
\, e^{i[\om(\pp)|t_1-t_2|+\pp(\x-\y)]}\,\, ,
\eeq
it follows that
\beq
2\int d\y \,\, G_0(\x_1,t_1,\y,t)\frac{\p}{\p t}G(\y,t,\x_2,t_2)=
\left\{ \ba{cr}
{\dis  G_0(\x_1,t_1,\x_2,t_2)} & t_1>t\,\, , t>t_2
\\
{\dis  -G_0(\x_1,t_1,\x_2,t_2)} & t_1<t\,\, , t<t_2
\\
{\dis  G_I(\x_1,t_1,\x_2,t_2)} & t_1<t\,\, , t>t_2
\\
{\dis  -G_I(\x_1,t_1,\x_2,t_2)} & t_1>t\,\, , t<t_2
\ea
\right.
\label{eq28}
\eeq
where $G_I$ is the ``image propagator'' equal to the free space
propagator for the points $(\x_1,t_1)$ and the reflection of $(\x_2,t_2)$
in the plane at time $=t$. In short, if the two points are on opposite
sides of the plane at time $=t$, the two propagators are ``glued''
to form the usual propagator, up to a sign, if they are on the same side the
gluing produces the image propagator. In the next section we will
interpret this relation in terms of the geometry of random paths. It
should not be confused with the self-reproducing property of heat-kernels,
but plays a nonetheless fundamental role in field theory. For example,
applying it twice leads to
\beq
\ba{c}
{\dis \int d\x_2\, d\x_3 \,\, G_0(\x_1,t_1,\x_2,t_2)
\bigg( \frac{-\p^2}{\p t_2\, \p t_3 }4G_0(\x_2,t_2,\x_3,t_3)\bigg)
G_0(\x_3,t_3,\x_4,t_4)=}\\
\\
{\dis = G_0(\x_1,t_1,\x_4,t_4) \,\,\,\,\,\,\,\,\,
 {\mbox for} \,\,\,\, t_1>t_2>t_3>t_4}
\ea
\eeq
\begin{figure}[htp]
\centering
\includegraphics[angle=0,width=0.65\textwidth]{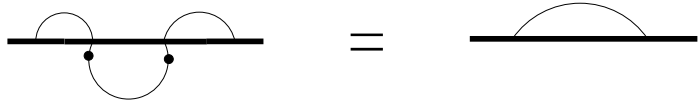}
\caption{The propagator at equal times is the inverse of $2\G_2$.}
\label{fig5}
\end{figure}
Taking all the $t_i$ to zero gives a relation which may be expressed
graphically as in figure \rf{fig5}, which shows that the
propagator at equal times is the inverse of $2\G_2$, with the
time-splitting regularization we discussed earlier.
\begin{figure}[htp]
\centering
\includegraphics[angle=0,width=0.25\textwidth]{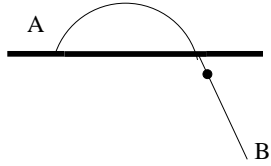}
\caption{External legs glued to propagators.}
\label{fig6}
\end{figure}
\begin{figure}[htp]
\centering
\includegraphics[angle=0,width=0.25\textwidth]{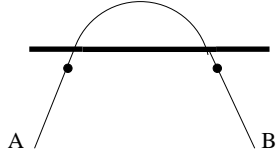}
\caption{``Gluing'' with two propagators.}
\label{fig7}
\end{figure}

Now consider a general term in the expansion of the two-point
function. The inverse of $(-2\G_2^1)$ appear either as external legs glued
to  propagators as shown in figure \rf{fig6}, or they appear glued
to two propagators as in figure \rf{fig7}. If $A, B$ are the points
$(\x,t_1)$ and $(\x_2,t_2)$ then the component in figure \rf{fig6} can
be evaluated using \rf{eq28} as the limit as $t_1$ and $t \rightarrow 0$
with $t_1>t$
\beq
\int d\y \,\, G_0(\x,t_1,\y,t) \frac{\p}{\p t}2G_0(\y,t,\x_2,t_2)=
G_0(\x,0,\x_2,t_2)
\eeq
i.e. gluing $(-2\G_2^1)^{-1}$ onto the Dirichlet propagator on the
boundary turns it into the free-space propagator restricted to the
boundary.  The second component, figure \rf{fig6}, is also simplified
using the gluing relation with $t_1<t<t'$, $t'>t_2$ and both $t,t'
\rightarrow 0$
\beq
\int d\y \,d\z \,\,\bigg( \frac{\p}{\p t}2G_0(\x,t_1,\y,t) \bigg)
 G_0(\y,t,\z,t') \frac{\p}{\p t'}2G_0(\z,t',\x_2,t_2)=
G_I(\x,t_1,\x_2,t_2)
\eeq
\begin{figure}[htp]
\centering
\includegraphics[angle=-90,width=0.75\textwidth]{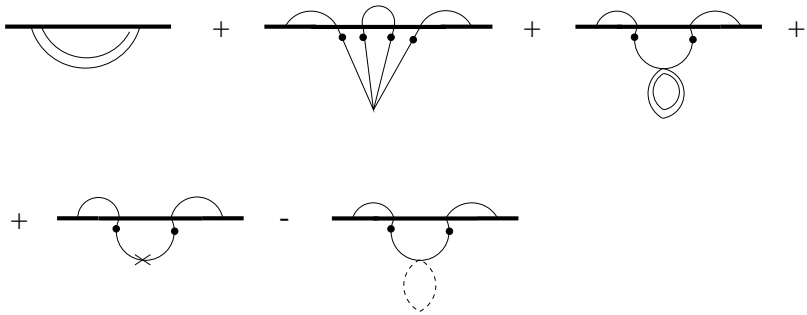}
\caption{The equal time two-point function.}
\label{fig8}
\end{figure}
\begin{figure}[htp]
\centering
\includegraphics[angle=-90,width=0.75\textwidth]{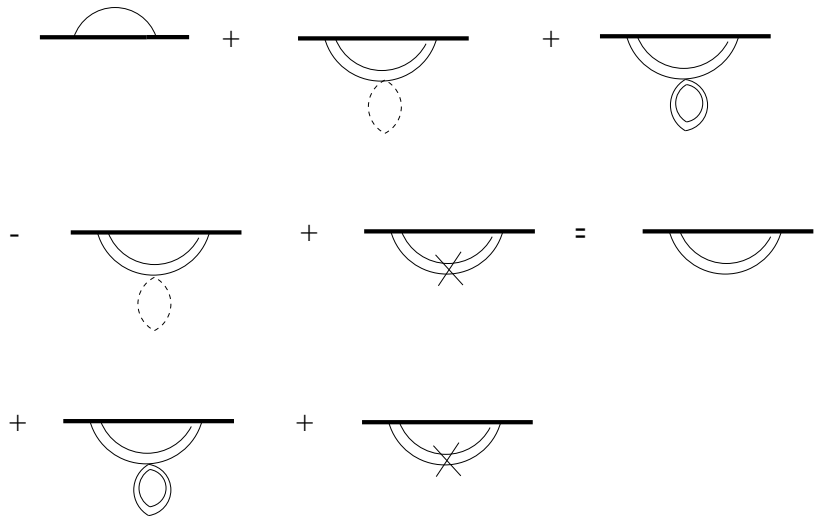}
\caption{Equal time two-point function once we have taken into account the
``gluings''.}
\label{fig9}
\end{figure}
\begin{figure}[htp]
\centering
\includegraphics[angle=0,width=0.4\textwidth]{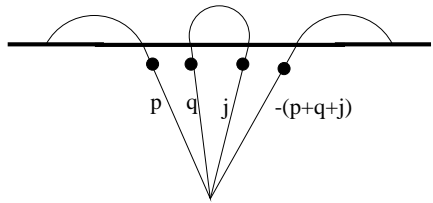}
\caption{Diagram to be cancelled.}
\label{fig10}
\end{figure}
So the effect of $(-2\G_2^1)^{-1}$ as an internal line in a diagram is to
produce an image propagator. This cancels against the image propagator
part of the Dirichlet propagator contributing from another diagram. So
if we denote the image propagator by a dotted line, (and the free-space
propagator by a double line, as before) then the equal time two-point
function is shown in figure \rf{fig8}, and with the above ``gluings''
we get the figure \rf{fig9}, which is just the figure \rf{fig4}  with
the free end-points restricted to $t=0$.  This cancellation may be
made explicit at the level of integrals where the diagram in figure
\rf{fig10} will be written as
\beq
\ba{c}
{\dis \int \frac{d\q\, dq_0\,dj_0 \, dp_0}{(2\pi)^4}\,\,
\frac{1}{\om(\pp)}\,ip_0\,\frac{1}{p_0^2+\om^2(\pp)}\,iq_0\,\frac{1}
{q_0^2+\om^2(\q)}\,\frac{1}{\om(\q)}\,ij_0}\\
\\
{\dis \frac{1}{j_0^2+\om^2(\q)}\,(-i(
p_0+q_0+j_0))\,\frac{1}{(p_0+q_0+j_0)^2+\om^2(\pp)}\,\frac{1}{\om(\pp)}}
\ea
\eeq
\begin{figure}[htp]
\centering
\includegraphics[angle=-90,width=0.4\textwidth]{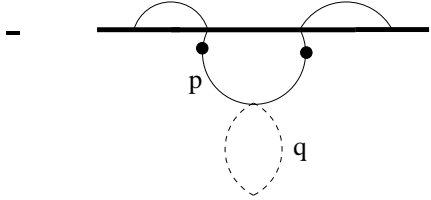}
\caption{Diagram with image loop (to be cancelled with the previous one).}
\label{fig11}
\end{figure}
Spatial momentum is conserved at each gluing, momentum and energy
are conserved at the four-point vertex. Whereas in the diagram in figure
\rf{fig11}, which is given by
\beq
-\int \frac{d\q\, dq_0\, dp_0}{(2\pi)^3}\,\,
\frac{1}{\om(\pp)}\,\frac{ip_0}{p_0^2+\om^2(\pp)}\,\frac{1}
{q_0^2+\om^2(\q)}\,\frac{(-i)(p_0+2q_0)}{(p_0+2q_0)^2+\om^2(\pp)}\,
\frac{1}{\om(\q)}\,,
\eeq
the image propagator causes energy not to be conserved at the vertex,
or rather to be ``conserved'' with a change of sign.

The two diagrams may be written as
\beq
D_1=\int d\q\, dq_0\,dj_0 \,\, \frac{q_0j_0\, f(q_0-j_0)}
{(q_0^2+\om^2(\q))(j_0^2+\om^2(\q))\om(\q)}
\eeq
and
\beq
D_2=-\int d\q\, dq_0\,\, \frac{f(2q_0)}{q_0^2+\om^2(\q)}
\eeq
with the same function $f$. If we change variables in $D_1$ from $q_0, j_0$
to $p_{\pm}\equiv\um(q_0\pm j_0)$ then
\beq
\ba{lcl}
D_1 & = & {\dis \int d\q\, dp_+\,dp_- \,\, \frac{(p_++p_-)(p_+-p_-)
\, f(2p_-)}{((p_++p_-)^2+\om^2(\q))((p_+-p_-)^2+\om^2(\q))\om(\q)}=}
\\
 &=& {\dis \int dp_- \, d\q\,\, \frac{f(2p_-)}{p_-^2+\om^2(\q)}\,\,=
\,\,-D_2\,\,\, ,}
\ea
\eeq
so that $D_1$ and $D_2$ cancel, as our general argument implied. Another way
to obtain the same result is by using \rf{deltaP}.

To end this section we will relate our diagrammatic method to the canonical
operator formalism. For that, we will use the formula (derived in appendix A)
\beq
\Psi_0[\phi]=\lim_{t\rightarrow\inf}e^{tE_0^{(i)}}\langle\,\phi(\x)\,|Te^{ -
\int_{-t}^0 dt'\,\, \hH_i(t')}|\,\Psi_0^{(0)}\rangle
\label{Dyson0}
\eeq
where $\hH_i(t)$ is the interaction Hamiltonian with the free
evolution and $\Psi_0^{(0)}$ the free VWF.  This equation has
the advantage that it can be easily related to the Rayleigh-\Sc
perturbation expansion (see Appendix A) and that it
gives the interaction-picture VEVs.
Previously, we have shown
that we could calculate the VWF if we introduced Dirichlet-propagators
and a new interaction term in the action. We will outline how
\rf{Dyson0} will give the same diagrammatic procedure.

We can derive the same boundary diagrammatics as before if we move
the $\phi$-dependence in \rf{Dyson0} to the interaction by using the
eq.  \rf{prod-exp}. Now we only need to consider the dynamics of free
fields with a Dirichlet boundary. Again we can implement this boundary
condition by using the method of image methods, with image charges
on the other side of the boundary:
\beq
\hf(x,t)=\frac{1}{\sqrt{2}}(\hf_0(x,t)-\hf_0(x,-t))
\eeq
where $\hf_0(x,t)$ is the field operator with free evolution and no
boundary. With this definition, the field operator vanishes at the
boundary and the propagator is just the previous Dirichlet
propagator \rf{D-prop}.  Expanding the exponential and using Wick's
theorem reduces \rf{Dyson0} to a combination of Dirichlet
propagators and boundary interactions term with the same
diagrammatic interpretation as before.

Now we can construct the VEVs using the equations
\rf{Dyson2} and its conjugate \rf{conj}
\beq
\ba{c}
{\dis \langle\,\Psi_0\,|\phi(\x_1,0)\cdot\cdot\cdot\phi(\x_n,0)|\,\Psi_0\,\rangle =
\lim_{t\rightarrow\inf}e^{2tE_0^{(i)}}}\\
\\
{\dis\langle\,\Psi_0\,^{(0)}|Te^{ - \int_0^t
dt'\,\, \hH_i(t')}\phi(\x_1,0)\cdot\cdot\cdot\phi(\x_n,0)\, Te^{ -
\int_{-t}^0 dt'\,\, \hH_i(t')}|\Psi_0^{(0)}\rangle= }\\
\\
{\dis =\lim_{t\rightarrow\inf}e^{2tE_0^{(i)}}
\langle\,\Psi_0\,^{(0)}|T\phi(\x_1,0)\cdot\cdot\cdot\phi(\x_n,0)\,e^{ -
\int_{-t}^t dt'\,\, \hH_i(t')}|\Psi_0^{(0)}\rangle}
\ea
\eeq
We recognize in the last equality the usual interaction-picture
formula which, by expanding the exponential, gives the usual Feynman
diagrams. The previous interpretation of
\rf{Dyson0} in terms of boundary diagrams implies that, when we
compute the VEVs, all these boundary diagrams combine to reproduce the usual
Feynman diagrams, as they should.


\section{Interpretation of the Vacuum functional in terms of random paths}
The Feynman diagram expansion of the VWF has a simple interpretation
in terms of the random paths of first quantisation allowing a geometric
understanding of the `gluing relation'\rf{eq28}.
It is well known that the Euclidean free-space propagator from
$A$ to $B$ may be written as a sum over all paths from $A$ to $B$ of
a Boltzman weight given by the exponential of the length of the  path.
 Explicitly this sum may be expressed in the path integral form
\beq
\int \D \x\,\, e^{-M\int_0^1 d\xi\,\, \sqrt{\dot\x\cdot\dot\x}}
\label{eq2.1}
\eeq
where $\xi$ parametrises the path and $\x(0)$ is the point $A$,
$\x(1)$ the point $B$. Alternatively this may be written in the form
\beq
\int \D \x\D e\,\, e^{-\int_0^1 d\xi\,\, (\dot\x\cdot\dot\x/e+Me)}
\label{eq2.2}
\eeq
where the square-root has been eliminated at the cost of introducing an
additional degree of freedom, $e$, which plays  the role of an
intrinsic metric. The Feynman diagram expansion of $\Psi$
needs the Dirichlet propagator rather than the free space one. We will
now show that this is given as a sum over paths on the space-time in
which points are identified with their reflection in the quantisation
surface, $t=0$, and paths are weighted with a minus sign every time
they cross this surface. Even though the path integrals \rf{eq2.1} and
\rf{eq2.2} are one-dimensional (i.e. quantum mechanical) they still
require regularization. This can be done by expanding $\x$ about a
classical solution to the Euler-Lagrange equations (satisfying the
boundary conditions that paths run from $A$ to $B$) as  a Fourier
sine-series truncated at some short wave-length
\beq
\x(\xi)=\x^{class} + \sum_{n=1}^N \x_n \sin(n\pi)
\eeq
\begin{figure}[htp]
\centering
\includegraphics[angle=-90,width=0.5\textwidth]{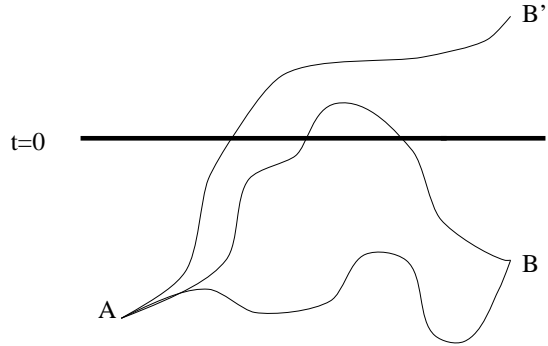}
\caption{Sum over paths from A to B and from A to B' (the reflection of
B).}
\label{fig12}
\end{figure}
This restricts us to differentiable paths. When we sum over paths from
$A$ to $B$ we have to include paths from $A$ to $B'$, the reflection of
$B$, since $B$ and $B'$ are considered equivalent (see figure
\rf{fig12}). Such paths cross the quantisation surface an odd number of
times and so acquire an overall minus sign, whereas paths directly from
$A$ to $B$ cross an even number of times and so are weighted with an
overall plus sign. The contribution from the latter paths gives $G_0$
and from the former $-G_I$ in expression \rf{D-prop} for the Dirichlet
propagator.
\begin{figure}[htp]
\centering
\includegraphics[angle=-90,width=0.5\textwidth]{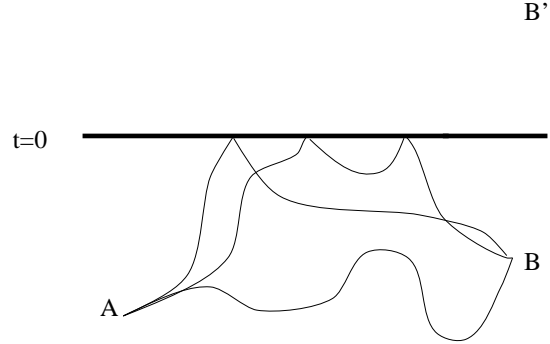}
\caption{Previous diagram where we identify each path that crosses $t=0$
with another path that is reflected at $t=0$.}
\label{fig13}
\end{figure}

Alternatively we can identify each path that crosses $t=0$ with another
path that is reflected at $t=0$ and so confined to $t<0$. For example
the paths in figure \rf{fig12} are identified with those in figure
\rf{fig13}. Now we attach a minus for each reflection. The paths have
the same lengths as previously and so  again lead  to the Dirichlet
propagator. There is no double counting because the reflected paths in
figure \rf{fig13} do not appear in the previous sum as they are not
differentiable at t=0. When this representation of the Dirichlet
propagator is inserted into the Feynman diagram expansion we arrive at
an expression for $W$ as a sum over {\it networks} of paths in $t<0$
that are reflected at the boundary.

Finally, the gluing property of the free-space propagator $G_0$:
$$ \int d\y\,\, G_0(\x_1,t_1,\y,t)2\frac{\p}{\p t}G_0(\y,t,\x_2,t_2)=
G_0(\x_1,t_1,\x_2,t_2)  \;\;\;\;\;\;\;\; \textrm{for}\;\;\;\;
t_1>t>t_2 $$
\begin{figure}[htp]
\centering
\includegraphics[angle=-90,width=0.5\textwidth]{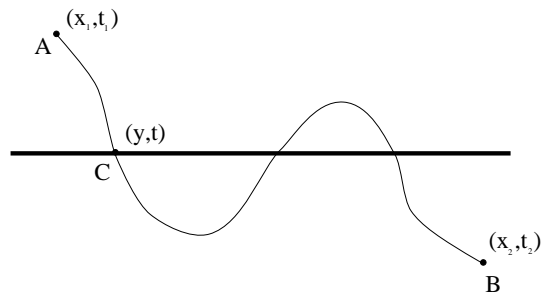}
\caption{Paths from $(\x_1,t_1)$ to $(\x_2,t_2)$
must cross the plane at time $t$ at least once.}
\label{fig14}
\end{figure}
simply reflects the fact that paths from $(\x_1,t_1)$ to $(\x_2,t_2)$
must cross the plane at time $t$ at least once (if $t_1>t>t_2$)
allowing the sum over such paths to be factorized (see figure
\rf{fig14}) so that formally
\beq
\ba{l}
{\dis \sum_{\textrm{paths AB}} e^{-\textrm{length(AB)}}=}\\
\\
{\dis  \sum_\y  \bigg( \sum_{\textrm{paths AC}}
e^{-\textrm{length(AC)}}\bigg) \bigg( \sum_{\textrm{paths CB}}
e^{-\textrm{length(CB)}}\bigg) }
\ea
\eeq
We note in passing that the first quantized path integrals \rf{eq2.1}
and \rf{eq2.2} have well-known generalizations to string theory
suggesting that our considerations in this section have relevance to
String Field Theory, and extend immediately to the Anti de Sitter space-time
considered in \cite{witten}.


\section{Analyticity of VEVs in the cut-off}

The purpose of this  paper is to show how VEVs can be reconstructed
from a derivative expansion of the VWF. This relies on a knowledge of
the domain of analyticity of VEVs in a momentum cut-off which we will
now discuss.  We begin with the representation of the VWF \rf{eq6} and
use the time splitting regularization for convenience. Our basic
assumption is that $\Psi_0[\f]$ can be expanded in positive powers of
$\f$ (we will give an argument to justify this), so
that
\beq
\Psi_0[\f]=\int \D\phi\,\, e^{-S_E[\phi]}\,\bigg(
\sum_{n=0}^{\inf} \frac{1}{n!}\prod_{i=1}^{n} \int dt\, d\x\,\,
\f(\x)\,\dot{\phi}(\x,t)\d_\ep(t)\bigg)
\label{eq3.1}
\eeq
We can now
reinterpret this Euclidean field theory on the space-time $t<0$ by
interchanging the roles of space and Euclidean time, so that we obtain
a theory on the space-time $x>0$. $\dot{\phi}(\x,t)$ is then
replaced by $\phi'(x,t)$ and the functional integral is interpreted
in the canonical formalism as
\beq
\Psi_0[\f]=\sum \frac{1}{n!(2\ep)^n}\int \prod dt_i\, dx_i\,\, \f(t_i)
\langle\,0_r\,|T\prod_{i=1}^{n}\hat{\phi}'(x_i,t_i)|\,0_r\,\rangle
\label{eq3.2}
\eeq
where $|\,0_r\rangle$ is the vacuum of the rotated theory and the integration
over $x$ is understood to be through the interval $[-\ep,0]$.
Therefore we can interpret $\Psi_0[\f]$ as the VEV of the
evolution operator between $t=-\inf$ and $t=\inf$ of a time
dependent Hamiltonian $H(\f(t))$  corresponding to the rotated theory
at $x>0$. The field $\f(t)$ will now play the role of a time dependent
coupling. Consider slowly varying configurations  for which the {\it
interaction} term ($\int dt\,\, \f(t)\, \phi'(0,t)$) is adiabatic,
then $|\,0_r\rangle$ (originally an eigenstate of $H(\f(-\inf))$) will remain
at the instantaneous ground state of $H(\f(t))$. As a result of this,
$\Psi_0[\f]$ is given by the exponential of
$-\int\limits_{-\inf}^{\inf}dt\,\, E_0(\f(t))$. The $E_0(\f(t))$ energy
of the instantaneous Hamiltonian ($t$ is fixed) is then approximately
given by the effective potential at its minimum. If we change the sign
of the {\it coupling} $\f(t)$, then we do not expect that the
interaction term can change the position of the vacuum ($\int dt\,\, \f
\phi' \ll H$ for large $\phi$). Therefore $E_0(\f)$ will be analytic
for small $\f$ and expandable, as we wanted to show. The adiabatic
approximation will be good for slowly varying $\f(t)$, roughly constant on the
time scale $\sim1/(E_1-E_0)$, with $E_1$ being the energy of
the first excitation.

The time integrals in \rf{eq3.2} can be done if we Fourier analyze the
sources $\f$, and use
\beq
\hat{\phi}'(\ep,t)=e^{t \hat{H}_r}\, \hat{\phi}'(\ep,0) \, e^{-t \hat{H}_r}
\label{eq3.3}
\eeq
so that (we have normal ordered the Hamiltonian and assumed an $i\ep$
prescription)
\beq
\ba{c}
{\dis \Psi_0[\f]=\sum_{n=0}^\inf \int \prod_{i=1}^n dk_i\,\, \tf(k_1)\cdots
\tf(k_n)\, \d(k_1+\cdots + k_n) \, }\\
\\
{\dis \times \langle\,0_r\,|\hat{\phi}'\,\frac{1}{
\hat{H}_r+i\sum_{i=1}^{n-1}k_i}\cdots \frac{1}{
\hat{H}_r+i(k_1+k_2)}\,\hat{\phi}'\, \frac{1}{
\hat{H}_r+i k_1}\, \hat{\phi}'|\,0_r\,\rangle}\\
\\
{\dis \equiv \sum_{n=0}^\inf \int \prod_{i=1}^n dk_i\,\, \tf(k_1)\cdots
\tf(k_n)\, \d(k_1+\cdots + k_n) \, G(k_1,\dots,k_n)}
\ea
\label{eq3.4}
\eeq
We now focus on the analyticity properties of each Green's function
$G(k_1,\dots,k_n)$ when we scale all the momenta by $s^{-1/2}$
\beq
\ba{l}
{\dis G_n(\frac{k_1}{\sqrt{s}},\dots,\frac{k_n}{\sqrt{s}})= }\\
\\
{\dis \langle\,0_r\,|\hat{\phi}'\,\frac{1}{
\hat{H}_r+i\sum_{i=1}^{n-1}k_i/\sqrt{s}}\cdots \frac{1}{
\hat{H}_r+i(k_1+k_2)/\sqrt{s}}\,\hat{\phi}'\, \frac{1}{
\hat{H}_r+i k_1/\sqrt{s}}\, \hat{\phi}'|\,0_r\,\rangle}
\ea
\label{eq3.5}
\eeq
In order to decompose this matrix element into simpler elements we will
insert the resolution of the identity in terms of a basis
of energy eigenstates. The sums over the energy eigenvalues converge, since the vacuum
functional itself is finite. Even in the higher dimensional case
when an infinite wave-function renormalisation
is
needed along with boundary counter-terms (which depend on $\f$ and are polynomial in the momenta)
to ensure this finiteness
these sums will continue to converge as the cut-off is removed.
If they did not, then in the presence of a regulator
these sums would be finite, but dominated by large energy
contributions for which the $k_i$ would be negligible,
so that
when the cut-off is removed the vacuum functional would have trivial dependence on the momenta. Because these sums converge their singularities
occur where the denominators vanish.
Since the eigenvalues of $\hat{H}_r$ are real, as are the
$k_i$, the singularities of our expression
lie on the negative real axis of the complex s-plane.

We can illustrate this analyticity in
perturbation theory. The
expansion of eq.  \rf{vacpert} (where the eq. \rf{pert} is its leading
term) gives the VWF to an arbitrary order.
We will scale the field momenta by a factor of $s^{1/2}$ and then we
perform a change of variables in the momenta integrations (scaling the
momenta by $s^{-1/2}$). To a given order of the perturbation expansion $\Psi_0[\phi]$ is given by
a finite sum of terms
containing energy denominators ($E_0^{(0)}-E_n^{(0)}=\sum \om_k$, with
$\om_k=\sqrt{k^2/s+m^2}$), expectation values (like
$\langle\,\Psi_n^{(0)}\,|\hH_{int}|\,\Psi_0^{(0)}\,\rangle\equiv H^I_{n0}$) and wave functionals
$\Psi_n^{(0)}[\phi;\kk_1,\cdots,\kk_n]$. In general we have
\beq
\Psi_0[\phi]=\Psi_0^{(0)}[\phi]+\sum_{k=1}^\inf \sum_{n_1\cdots n_k}
\Psi_{n_1}^{(0)}[\phi] \frac{H^I_{n_1n_2}H^I_{n_2n_3}\cdots H^I_{n_k0}}
{(E^{(0)}_0-E^{(0)}_{n_1}-i\ep)\cdots (E^{(0)}_0-E^{(0)}_{n_k}-i\ep)}
\eeq
Where the $n_i$ sums also involve the momenta integrations.
The energy denominators have
the expected analytic behavior: analytic in the whole complex
$s$-plane with a cut from $s=-\L^2/m^2$ (the $\om_k$'s are integrated
up to the cut-off $\L$) to $s=0$.  The matrix
elements will also be a combination of $\om_k$ because they are
evaluated with Wick's theorem (remember that the creation operator is
given by $a^\da(k)=\int dx\,\, e^{-ikx}(\om_k\,\phi(x)-\d/\d\phi(x)$)
and therefore they have the same analytic behavior. Finally the wave
functionals, which can be constructed operating $a^\da(k)$ several
times into the free VWF (which can be written as $N\, e^{-\um\int
\frac{dk}{2\pi}\om_k\tilde\phi(k)\tilde\phi(-k)}$), are also analytic.
Therefore we obtain that the scaled VWF is analytic in the whole
complex $s$-plane with a cut from $s=-\L^2/m^2$ to $s=0$.

Relying also on the assumption that
we can treat $\Psi_0$ as a power series in $\f$ allows us to take
the logarithm of $\Psi_0$  to obtain the $\G_{2n}$ of \rf{eq17}
as sums of products of $G_m$ with $m\le 2n$ so that $\G_{2n}(
p_1/\sqrt{s},\dots,p_{2n}/\sqrt{s})$ is also analytic in this domain.

Now consider evaluating the Fourier transform of the equal time VEV
\rf{corr} by expanding all but the terms quadratic  in $\f$ in
$|\Psi_0|^2=\exp(2W)$:
\beq
\ba{c}
{\dis \int \D \f\,\, |\Psi_0[\f]|^2\, \tf(p_1)\cdots\tf(p_n)=}\\
\\
{\dis \int \D\f \,\, e^{\frac{2}{\hbar}\int dp\,\,\tf(p)\G_2(p,-p)\tf(p)}
\sum \frac{(2 \tilde{W}[\f])^n}{n!}\, \tf(p_1)\cdots\tf(p_n)}
\ea
\label{eq3.6}
\eeq
The expansion is similar to that of \rf{eq22}, but we keep contributions
to $\G_2$ of all orders in $\hbar$ in the exponent, and
$\tilde{W}$ consists of the remaining terms in $W$. Integrating over
$\tf$ with a momentum cut-off $p^2<\L$ leads to a sum of terms
in which we make contractions with $(\frac{2}{\hbar}\G_2)^{-1}$, so
if we write a typical term in the expansion of $\tilde{W}^n$ as
\beq
\int \prod_i dp_i \,\, \tf(p_i)\, H(p_1,\dots,p_n)
\eeq
then these contractions lead to sums of terms like
\beq
\ba{c}
{\dis \int_{q^2<\L}\prod_{i=1}^m dq_i\,\, H(q_1,-q_1,q_2,-q_2,
\dots ,q_m,-q_m,\dots ,p_1,\dots ,p_n) \cdot}\\
\\
{\dis \prod_{j=1}^m \G_2^{-1}(q_j,-q_j) \, \prod_{k=1}^n \G_2^{-1}(p_k,
-p_k)=K(p_1,\dots,p_n,\L)}
\ea
\eeq
In the $1+1$ dimensions we have been working in this is finite
as $\L\rightarrow \inf$. We want to show that it is computable from
a knowledge of an expansion of $W$ in positive powers of
momentum. Such an expansion would appear to be convergent only for
small momenta, if at all, so it would not appear to be useful for the limit
$\L\rightarrow\inf$. However analyticity in $\L$ allows us to
calculate large $\L$ behavior from small $\L$ behavior using Cauchy's
theorem. Consider the effect of scaling $\L$ by $1/s$ and the momenta
by $1/\sqrt{s}$
\beq
\ba{c}
{\dis K\Big(\frac{p_1}{\sqrt{s}},\dots,\frac{p_n}
{\sqrt{s}},\frac{\L}{s}\Big)=}\\
\\
{\dis \int_{q^2<\L}\prod_{i=1}^m \frac{dq_i}{\sqrt{s}}
\,\, H\Big(\frac{q_1}{\sqrt{s}},\frac{-q_1}{\sqrt{s}},
\dots ,\frac{q_m}{\sqrt{s}},\frac{-q_m}{\sqrt{s}}
,\dots ,\frac{p_1}{\sqrt{s}},\dots ,\frac{p_n}{\sqrt{s}}\Big) \cdot}\\
\\
{\dis \prod_{j=1}^m \G_2^{-1}\Big(\frac{q_j}{\sqrt{s}},\frac{q_j}
{\sqrt{s}}\Big) \, \prod_{k=1}^n \G_2^{-1}\Big(\frac{p_k}{\sqrt{s}},
\frac{-p_k}{\sqrt{s}}\Big)}
\ea
\label{eq3.8}
\eeq
For $s$ large all the arguments of $H$ and $\G_2$ are small so we can use a
small momentum expansion for these quantities. We have already
shown that $H$ and $\G_2$ are analytic in the complex $s$-plane cut
along the negative real axis so it follows that $K$ is analytic
in this cut-plane provided that $\G_2^{-1}$ is also analytic. This last
 result follows if the only zeroes of $\G_2(\frac{p}{\sqrt{s}},
\frac{-p}{\sqrt{s}})$ lie on the negative real axis. Since $\G_2(p,-p)$
is even in $p$, we obtain from \rf{eq3.4}
\beq
\G_2\Big(\frac{p}{\sqrt{s}},
\frac{-p}{\sqrt{s}}\Big)=\langle\,0_r\,|\hat{\phi}'\, \frac{\hat{H}_r}{
\hat{H}_r+\frac{p^2}{s}}\, \hat{\phi}'|\,0_r\,\rangle
\eeq
Inserting a basis of eigenstates of $\hat{H}_r$ gives
\beq
\G_2\Big(\frac{p}{\sqrt{s}},
\frac{-p}{\sqrt{s}}\Big)=\sum_{\ve}|\langle\,0_r\,|\hat{\phi}'|\,\ve\,\rangle|^2\,\frac{
\ve}{\ve+\frac{p^2}{s}}
\eeq
The imaginary part of this is
\beq
\Im \G_2\Big(\frac{p}{\sqrt{s}},
\frac{-p}{\sqrt{s}}\Big)=\bigg( \sum_{\ve}\frac{
|\langle\,0_r\,|\hat{\phi}'|\,\ve\,\rangle|^2\,\ve p^2}{\ve^2+\frac{p^4}{|s|^2}}
\bigg) \, \Im \big(\frac{1}{s}\big)
\eeq
Each term in the sum is greater than or equal to zero, and some
terms must be non-zero, else $ \hat{\phi}'|\,0_r\,\rangle=0$. So the imaginary
part of $\G_2$ can only be zero for finite $s$ when $s$ is real. The
real part of $\G_2$ is, for real s
\beq
\Re \G_2\Big(\frac{p}{\sqrt{s}},
\frac{-p}{\sqrt{s}}\Big)=\sum_{\ve}\frac{
|\langle\,0_r\,|\hat{\phi}'|\,\ve\,\rangle|^2}{\ve^2+\frac{p^4}{s^2}}
 \,  \big(\ve+\frac{p^2}{s}\big)
\eeq
which is a sum of positive terms for $s>0$ and hence can only vanish
when $s<0$. Thus the zeroes of $\G_2\Big(\frac{p}{\sqrt{s}},
\frac{-p}{\sqrt{s}}\Big)$ lie on the negative real axis and so
 \rf{eq3.8} is analytic in the cut $s$-plane. (A boundary counter term such as $\Lambda \f^2$ cannot spoil this,
because, if present, it affects only the momentum independent part of
$\Gamma_2$ cancelling any divergence.)
Furthermore, for large
 $s$ we expect that
$\G_2\Big(\frac{q}{\sqrt{s}},\frac{-q}{\sqrt{s}}\Big)$ and
$H\Big(\frac{q_1}{\sqrt{s}},\dots,\frac{q_n}
{\sqrt{s}}\Big)$ have a finite limit, so that \rf{eq3.8} behaves
like $s^{-m/2}$.
We conclude that the VEV with scaled momenta
and cut-off is a sum of terms each of which is analytic in the cut
$s$-plane (plus a neighborhood of $s=\inf$)
and drops off like a negative power of $s$ for large $s$.

Let $\tilde{K}$ be the sum of such contributions and define the contour
integral
\beq
I(\l)=\frac{1}{2\pi i}\int_C ds\,\, \frac{e^{\l(s-1)}}{s-1}\,
\tilde{K}\Big(\frac{p_1}{\sqrt{s}},\dots,\frac{p_n}
{\sqrt{s}},\frac{\L}{\sqrt{s}}\Big)
\label{resum}
\eeq
where $C$ is the circle at infinity starting below the cut on the
negative real axis and ending above it. On $C$ all the momenta in
\rf{eq3.8} are small and we can use the small momentum expansions for $H$
and $\G$, which leads to a power series in $1/\sqrt{s}$. Since $s$
is large we can rewrite this as a power series in $1/(s-1)$
\beq
\tilde{K}\Big(\frac{p_1}{\sqrt{s}},\dots,\frac{p_n}
{\sqrt{s}},\frac{\L}{\sqrt{s}}\Big)\, \sim
\sum A_n (p_1,\dots,p_n,\L)\,\frac{1}{(s-1)^n}\,\,
\,\,\,\,\,\,\,\,\,\,(s\rightarrow \inf )
\label{s-1}
\eeq
and the analyticity around $s=\inf$ implies that the coefficients
$A_n$ will grow as $C^n$ ($C$ is a constant). Because the
convergence radius around $s=\inf$ is proportional to the cut-off we
expect that these coefficients will grow with the cut-off (and if the
cut-off is eliminated then the $C^n$ behavior at large $n$ is spoiled).
Now
\beq
I(\l)=\sum A_n (p_1,\dots,p_n,\L)\,\frac{\l^n}{n!}
\label{Ilam}
\eeq
is the Borel transform of \rf{s-1}. This series, due to the $n!$,
will be convergent for all $\l$ (thanks to the previous $C^n$ growth
which was the consequence of the analyticity around $s=\inf$). If we
now collapse the contour we obtain a contribution from the pole at
$s=1$ which is the VEV we seek, $\tilde{K}(p_1,\dots,p_n,\L)$, together
with a contribution from the cut which is suppressed by the exponential
factor $e^{\l(s-1)}$ as $\l\rightarrow\inf$ (given that the
discontinuity in $\tilde{K}$ across the cut goes to zero as $s\rightarrow
-\inf$ as a power of $s$). Thus we recover the VEV as the limit
of the Borel transform
\beq
\tilde{K}(p_1,\dots,p_n,\L)=\lim_{\l\rightarrow\inf} I(\l)=
\lim_{\l\rightarrow\inf} \sum A_n (p_1,\dots,p_n,\L)\,\frac{\l^n}{n!}
\label{recov}
\eeq
And the coefficients $A_n$ are obtainable from the local
expansion of the VWF. We can also see from the integral \rf{resum} why
$I(\l)$ is finite: the analyticity around $s=\inf$ allows us to change
the integration contour in such a way that it will have a finite length
and it will not cross any singularity, and thanks to the exponential
suppression the limit $\lim_{\l\rightarrow\inf} I(\l)$ exists and is
finite. This implies that this series is alternating and, as we have
seen previously, also convergent for all $\l$. Although we need to
evaluate the power series for large $\l$, we can obtain the value of
the limit $\l\rightarrow\inf$ in the same way that we can estimate the
value of the limit $\lim_{x\rightarrow\inf} e^{-x}$ from the $x$
behavior of $\sum_{n=0}^N (-1)^n x^n/n!$ for $N$ as large as
$x^N/N!\ll 1$ (if we take $N=10$ then for $x<4$ this series gives
$e^{-x}$ with an error of $O(0.1)$ and therefore our estimate for
$\lim_{x\rightarrow\inf} e^{-x}$ will be $\sum_{n=0}^{10} (-1)^n
4^n/n!=0.097$ which is zero up to $O(0.1)$).

This leads to a controllable approximation in which  we truncate the
series at some order in $\l$. The error is then bounded by the
highest order term retained which tells us how large we can take $\l$.
How close this truncated series comes to displaying
the limiting behavior for large $\l$ can then be
judged from how flat the truncated series is as a function of $\l$
in the region of the largest value for which the truncation is
trustworthy.

  Before ending this section, it is interesting to see the connection
of our resummation technique and the usual dispersion relation method.
Consider the equation \rf{resum}, rescale $s$ to $s/\l$ and consider
the contour to go from $s=-\inf-i\ep$ to $s=-i\ep$, then to
$s=i\ep$ and finally to $s=-\inf+i\ep$ (we have separated the pole
term).  We get (to simplify, we just write the $s$-dependence)
\beq
I(\l)=\tilde{K}(1)+\frac{e^{-\l}}{2\pi i}
\int_{0}^\inf ds\,\, \frac{e^{-s}}{s+\l} \Delta\tilde{K}(s)
\label{disp}
\eeq
where $\Delta\tilde{K}(s)=\tilde{K}(-s/\l+i\ep)-\tilde{K}(-s/\l-i\ep)$
is the discontinuity across the cut and we have assumed an
infinitesimal $\ep$.  Therefore, for large $\l$, $\tilde{K}(1)$ is
determined by the low momentum expansion which will give us $I(\l)$,
and by an ``exponentially subtracted'' dispersive integral in contrast
to the usual subtractions by polynomials.
We also see that in the limit $\l\rightarrow\inf$
the integral in \rf{disp} only goes to zero as a power of $1/\l$,
because we have assumed that $\Delta\tilde{K}(s)$ was polynomially
bounded, and therefore we get the expected result \rf{recov}.


\section{Matrix elements in Quantum Mechanics by cut-off resummation}
In this section we will apply our method to
a one-dimensional non-relativistic quantum mechanical bound state
problem. Consider a Hamiltonian such that its ground state
wave-function is short ranged, vanishing
at least exponentially for large distances, and non-singular. This
condition implies that its Fourier transform $\tilde{\Psi}_0(k)$
\beq
\tilde{\Psi}_0(k)=\int\limits_{-\inf}^{\inf} dx\,\, e^{-ikx}\Psi_0(x)
\eeq
is analytic for $|k|<1/R_0$ if $\Psi_0(x)\sim e^{-|x|/R_0}$ for large
$|x|$ because the integral, and all its derivatives in $x$, are
convergent.  We can say more about the analyticity of
$\tilde{\Psi}_0(k)$, it is analytic in the whole complex plane with
poles at $k=i/R_0$ due to the upper integration limit and at
$k=-i/R_0$ due to the $x\rightarrow -\inf$ limit. If $\Psi_0(x)$
decays faster than that at large distances then $R_0\rightarrow\inf$.
It is enough for our purposes to consider systems with a short range
potentials and then $R_0$ will be finite.  Now we will introduce the
projector
\beq
\P\equiv\int_{|k|<1/\sqrt{s}}|\,k\rangle\langle\,k\,|
\eeq
to eliminate the degrees of freedom $k>1/\sqrt{s}$ in the computation of
matrix
elements. We refer the reader to the appendix B for some examples and
mathematical
details. Let us study the analyticity of a simple expectation value
\beq
E(s)=\langle\,\Psi_0\,|\P_{\frac{1}{\sqrt{s}}}\hat p^2
\P_{\frac{1}{\sqrt{s}}}|\,\Psi_0\,\rangle=
\int\limits_{|k|<\frac{1}{\sqrt{s}}}\!\!  dk \,\, k^2
|\tilde{\Psi}_0(k)|^2=\frac{1}{s^{3/2}} \int\limits_{|k|<1}\!\!
dk\,\, k^2 |\tilde{\Psi}_0(\frac{k}{\sqrt{s}})|^2
\label{p2v}
\eeq
the $|\tilde{\Psi}_0(\frac{k}{\sqrt{s}})|^2$ has poles at
$\sqrt{s}=\pm ikR_0$. Then the integral is analytic in the complex
$s$-plane with a cut in the negative real axis. We see that $s^{3/2}
 E(s)$ has the cut from $s=-R_0$ to $s=0$.  Notice that
in field theory, this projector would correspond to one in the Fourier
{\it field amplitudes} which is not the one we have
used before (where we have
restricted the Fourier components), but the main purpose of a
regulator is to discard some kind of degree of freedom and we do not
consider that our method depends on the form of the regulator. In fact,
in field theory will be sometimes more useful to use some {\it soft}
cut-off regulators which will preserve some wanted symmetries of the
theory. More generally, we can consider
\beq
\langle\,\Psi_n\,|\P_{\frac{1}{\sqrt{s}}}\hat A
\P_{\frac{1}{\sqrt{s}}}|\,\Psi_0\,\rangle=
\frac{1}{s}\int\limits_{|k,k'|<1}\!\!\! dk\, dk' \,\,
\tilde{\Psi}_n^*(\frac{k}{\sqrt{s}})
\tilde{\Psi}_m(\frac{k}{\sqrt{s}}) \, \langle\,\frac{k}{\sqrt{s}}|\hat
A|\,\frac{k'}{\sqrt{s}}\rangle
\label{Av}
\eeq
and assume that
$\langle\,x|\hat A|\,x'\rangle$ vanishes, when $|x-x'|\rightarrow\inf$, at least as
$e^{-|x-x'|/R}$ (in particular, that $\langle\,k\,|\hat A|\,k'\rangle$ is proportional
to $\d(k-k')$ and has poles for $k=\pm i/R$ with non-zero $R$). For
such local operators we can perform both integrals and
we get analyticity in the $s$-plane with a cut in the negative real
axis (again, sometimes we may multiply the matrix element by some power
of $s$ to get an analytic function in $s$ with a cut from $s=-\max(
R_0, R)$ to $s=0$. Alternatively we could have said that because the
wave functions and $\langle\,x\,|\hat A|\,x'\,\rangle$ decrease exponentially we obtain
convergent integrals even if we take an arbitrary number of derivatives
of them. In field theory we were only able to study the analyticity by
expanding the wave functions (and only considering ultralocal
functional A operators) because we had still unbounded integrals ($\int
d\tilde\phi_k$ integrals) which can change the analyticity domain if
these integrals diverge.

Now, suppose that we got the ground state wave
function for small $k$, then we expand it around $k=0$ (this is the
analogous to the derivative expansion of the VWF) and use the resummation
method to compute a matrix element like \rf{p2v} (we will use, as cut-off,
$\L$ for simplicity, in the appendix B it is explained the difference with
$1/\sqrt{s}$)
\beq
\int\limits_{|k|<\L}\!\!  dk \,\, k^2
|\tilde{\Psi}_0(k)|^2=\int\limits_{|k|<\L}\!\!  dk \,\, k^2
(a_0-k^2a_2+\cdots)=(\L^3\frac{a_0}{3}-\L^5
\frac{a_2}{5}+\cdots)
\label{k2gsL}
\eeq
if we only consider these two terms, the resummed value will be estimated by
the value of $\L^3\frac{a_0}{3\cdot 3!}-\L^5
\frac{a_2}{5\cdot 5!}$ at its local maximum, which gives
$4 a_0^{5/2}/a_2^{3/2}$.  As an example, we will
consider a $\d(x)$, atractive, potential. Then
\beqr
\Psi_0(x)&=&e^{-\a |x|}\nn\\
\tilde{\Psi}_0(k)&=&\frac{2\a}{\a^2+k^2}
\label{Dgs}
\eeqr
and
\beq
|\tilde{\Psi}_0(k)|^2=\frac{4}{\a^2}-\frac{8}{\a^4}k^2+\frac{12}{\a^6}
k^4+\cdots
\eeq
In the appendix B we discuss the details of the resummation of \rf{k2gsL}
and of $\langle\,\Psi_0\,|\,\Psi_0\,\rangle$. Although the quantum mechanical case is trivial it
is useful as a test ground. Notice that we do not attempt to construct the
local expansion coefficients (i.e. $\tilde{\Psi}_0(k)$ at small $k$) from
a given Hamiltonian. We only give a method to compute amplitudes once a
wave-function at large distances is given.


\section{Effective expansion}
In this section we will describe how the computation of VEVs with a given
local expansion of the VWF automatically leads to a new ``infrared'' diagramatic
approach. The VEVs will arise a the result of the resummation (i.e. we will
find the analytic continuation of the series for the region of large cut-off
in a systematic way).
Suppose we know the leading terms of the local expansion of $W[\tf]$
\beqr
2 W[\tf]&=&\frac{-1}{2}\int \frac{dp}{2\pi}\,\,\tf(p)\tf(-p)\,\,
(\a_0+\a_2 p^2+\cdots)-\nn\\
&&-\frac{1}{4!}\int \frac{dp_1}{2\pi}\cdots \int
\frac{dp_4}{2\pi}2\pi\d(\sum p_i) \tf(p_1)\cdots\tf(p_4)\nn\\
&&\times (\b_0+\b_2(p_1^2+
\cdots +p_4^2)+\cdots)+O(\tf^6)
\eeqr
\begin{figure}[htp]
\centering
\includegraphics[angle=0,width=0.85\textwidth]{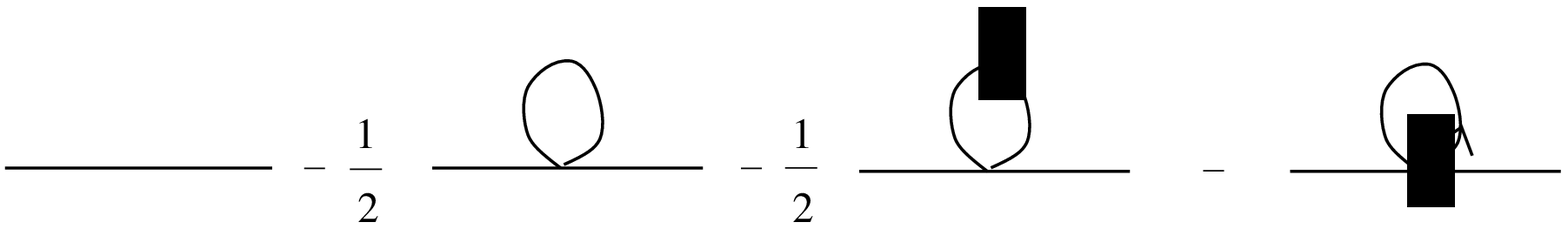}
\caption{Two-point Green's function for zero momenta at $O(\b\L^3)$.}
\label{fig15}
\end{figure}
where we have discarded the $O(p^4)$ terms. Having this information at
hand, we want to compute an equal-time connected Green's function. For
simplicity, we choose the two-point correlation function, which we compute
using eq. \rf{eq3.6}. We have to perform a perturbation expansion with a
$\f^4$ model in $1+1$ dimensions, with derivative interactions and an
explicit small cut-off $\L$. We will obtain our correlator as a series in
$\L$, $p$. Because we have truncated $W[\tf]$ we will discard the terms
$\L O(\L^4,p^4,\L^2 p^2)$. In figure \rf{fig15} we show the diagrams
corresponding to the contribution to the correlator at zero momenta. Due to
the fact that the cut-off is small, we consider the $\a_2$ term as a
$(\p\f)^2$ insertion (which we represent by a dot) and our
(momentum-independent) propagator will be $1/\a_0$. The vertex with a dot is
a $\b_2$ vertex, with the arrow indicating where the momenta are sitting.
Then we respectively get for each diagram:
\beq
\frac{1}{\a_0}-\frac{1}{2}\frac{\b_0}{\a_0^2}
\int\limits_{|q|<\L}\frac{dq}{2\pi}\frac{1}{\a_0}-
\frac{1}{2}\frac{\b_0}{\a_0^2}
\int\limits_{|q|<\L}\frac{dq}{2\pi}\frac{1}{\a_0}(-\a_2 q^2)\frac{1}{\a_0}
-\frac{\b_2}{\a_0^2}
\int\limits_{|q|<\L}\frac{dq}{2\pi}\frac{q^2}{\a_0}
\eeq
which gives
\beq
\frac{1}{\a_0}-\frac{1}{2\pi\a_0}\frac{\b_0}{\a_0}\frac{\L}{\a_0}+
\frac{1}{6\pi\a_0}\frac{\b_0}{\a_0}\frac{\L^3}{\a_0^2}\a_2-
\frac{1}{3\pi\a_0}\b_2\frac{\L^3}{\a_0^2}.
\label{tad0}
\eeq
What are the dimensionless expansion parameters? To answer that, we scale
the field $\f$ to absorb the $\a_2$  coefficient and, at the same time, we
note that (because we have a super-renormalizable model) the perturbative
expansion is performed on the coupling divided by the mass term. Therefore,
we will get as dimensionless couplings
$\tilde\b_0=\frac{\b_0}{\a_0}\frac{1}{\sqrt{\a_0\a_2}}$ and
$\tilde\b_2=\frac{\b_2}{\a_2}\frac{1}{\sqrt{\a_0\a_2}}$, with $\tilde\L=
\frac{\L}{\sqrt{\a_0/\a_2}}$ as a cut-off. Then we may rewrite \rf{tad0} as
\beq
\frac{1}{\a_0}(1-\frac{1}{2\pi}\tilde\b_0 \tilde\L +\frac{1}{6\pi}\tilde\b_0
\tilde\L^3-\frac{1}{3\pi}\tilde\b_2 \tilde\L^3)
\eeq
We assume that $\tilde\b_0$, $\tilde\b_2 \ll 1$. Our approximation to the
$\tilde\L\ra\inf$ limit is obtained by applying \rf{recov} to the
resummation of the $\tilde\L$-series (see appendix B). Although we have very
few terms, we can still do the resumation. Because the series has a finite
limit for $\tilde\L\ra\inf$ our best estimate will be a stationary value
(see also the remarks
given in the appendix B).

Applying this to the local expansion of the
perturbative $W[\f]$ to $O(g)$ as given by eq. \rf{Gtree} and \rf{G2},
we have
\beqr
\G_2(p,-p)&=&\frac{-1}{8\pi}(\a_0+\a_2 p^2+\cdots)=\nn\\
&=& \frac{-\om(p)}{4\pi}+\frac{g}{32\pi}\int\frac{dq}{2\pi}\frac{1}
{\om(q)(\om(q)+\om(p))}-\frac{\d M^2}{8\pi \om(p)}\\
\G_4(p_1,\ldots,p_4)&=&\frac{-1}{2(2\pi)^3}\frac{1}{4!}
(\b_0+\b_2(p_1^2+\cdots +p_4^2)+\cdots)=\nn\\
&=& \frac{-g}{(2\pi)^3 4!(\om(p_1)+\cdots +\om(p_4))}
\eeqr
using $\d M^2=g/(4\pi)$, we obtain the local expansion coefficients by
expanding the right-hand sides in powers of $p$, getting
\beqr
\a_0 &=& 2m \nn\\
\a_2 &=& (m-\frac{m}{12\pi}\frac{g}{m^2})\frac{1}{m^2}
\label{alphas}
\eeqr
and
\beqr
\b_0 &=& \frac{m}{2}\frac{g}{m^2}\nn\\
\b_2 &=& -\frac{m}{16}\frac{g}{m^2}\frac{1}{m^2}
\label{betas}
\eeqr
when we substitute these values into the zero momentum correlator (eq.
\rf{tad0}) we get
\beq
\frac{1}{2m}-\frac{1}{32\pi m}\frac{g}{m^2}\frac{\L}{m}+
\frac{1}{128\pi m}\frac{g}{m^2}\frac{\L^3}{m^3}+\frac{1}{m}
O(\frac{g^2}{m^4},\frac{\L^5}{m^5})
\label{c0}
\eeq
\begin{figure}[htp]
\centering
\includegraphics[angle=0,width=0.6\textwidth]{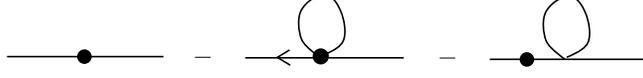}
\caption{$p^2$ contribution to the two-point Green's function at $O(\b\L p^2)$.}
\label{fig16}
\end{figure}
We can also compute the $p^2$ term of the two-point correlator. This is
given by the diagrams of the figure \rf{fig16}, where we have neglected
the $O(\L^2 p^2/m^4)$ terms, from which we get
\beq
-\frac{\a_2}{\a_0^2}p^2-\frac{p^2}{\a_0^2}\b_2
\int\limits_{|q|<\L}\frac{dq}{2\pi}\frac{1}{\a_0} +
p^2 \frac{\a_2 \b_0}{\a_0^3}
\int\limits_{|q|<\L}\frac{dq}{2\pi}\frac{1}{\a_0}
\eeq
which, after the substitutions \rf{alphas} and \rf{betas}, gives
\beq
\left( \frac{-1}{4m}+\frac{1}{48\pi m}\frac{g}{m^2}+\frac{5}{128\pi m}
\frac{g}{m^2}\frac{\L}{m}\right) \frac{p^2}{m^2}+O(\frac{g^2}{m^4})
\label{cp2}
\eeq
if we want to get right the $g \L^3 p^2/m^7$ term, we have to include
(in addition to the vertices already present in the action, but omitted
in figure \rf{fig16}) $\b_4$ vertices which will also give $q^2 p^2$
terms. Notice that we only got a $O(\L/m)$ contribution in \rf{cp2}, in
contrast with the $O(\L^3/m^3)$ of \rf{c0}.

The resummation of \rf{c0} gives
\beq
\frac{1}{2m}-\frac{1.88}{32\pi m}\frac{g}{m^2}
\label{c0res}
\eeq
we can check these results by computing our connected correlation
function without locally expanding the $W[\tf]$ term
\beqr
\int \D\tf\,\, \tf(p)\tf(-p)\, e^{2 W[\tf]}&=&
\frac{1}{2\tilde\G_2(p,-p)}-\um \left(
\frac{1}{2\sqrt{p^2+m^2}}\right)^2\cdot \nn\\
&&\cdot\int\limits_{|q|<\L}\frac{dq}{2\pi}\,\,\frac{1}
{2\tilde\G_2(q,-q)}\,
\frac{g}{\om(q)+\om(p)}+O(g^2)
\label{cnl}
\eeqr
where $\tilde\G_2=-4\pi \G_2$, then $2\tilde\G_2=\a_0+\a_2 p^2+\cdots$.
The second term of the right-hand side is the tadpole diagram with
dressed propagator and vertex (we have a non-local $\f^4$ vertex). The
propagator is
\beqr
\frac{1}{2\tilde\G_2(p,-p)}&=& \frac{1}{2\sqrt{p^2+m^2}} +
\frac{g}{16}\left(
\frac{1}{\sqrt{p^2+m^2}}\right)^2\int\frac{dq}{2\pi}\,\frac{1}
{\om(q)(\om(q)+\om(p))}-\nn\\
&& -\frac{g}{16\pi}\left( \frac{1}{\sqrt{p^2+m^2}}\right)^2
\frac{1}{\om(p)} + O(g^2)
\eeqr
then we get for \rf{cnl}
\beq
\ba{c}
{\dis \frac{1}{2\sqrt{p^2+m^2}}+\frac{g}{16}\left(
\frac{1}{\sqrt{p^2+m^2}}\right)^2\int\frac{dq}{2\pi}\,\frac{1}
{\om(q)(\om(q)+\om(p))}-}\\
\\
{\dis -\frac{g}{16\pi}\left(
\frac{1}{\sqrt{p^2+m^2}}\right)^2 \frac{1}{\om(p)} -\frac{g}{16}\left(
\frac{1}{\sqrt{p^2+m^2}}\right)^2\int\limits_{|q|<\L}
\frac{dq}{2\pi}\,\frac{1}
{\om(q)(\om(q)+\om(p))} + O(g^2)}
\label{cnl2}
\ea
\eeq
We realize that when $\L\ra\inf$, the second and fourth term cancel.
This is precisely the  cancellation between the diagram of figure
\rf{fig11} with the one of figure \rf{fig10} respectively. Now, we
expand \rf{cnl2} in powers of $\L/m,p/m$
\beq
\ba{l}
{\dis\left( \frac{1}{2m}-\frac{1}{32\pi m}\frac{g}{m^2}\frac{\L}{m}+
\frac{1}{128\pi m}\frac{g}{m^2}\frac{\L^3}{m^3} \right) +}\\
\\
{\dis + \left( \frac{-1}{4m}+\frac{1}{48\pi
m}\frac{g}{m^2}+\frac{5}{128\pi m} \frac{g}{m^2}\frac{\L}{m}\right)
\frac{p^2}{m^2}+O(\frac{p^4}{m^4})}
\ea
\eeq
we have dropped the terms $\L^3/m^3,p^2/m^2$. If instead of expanding
in the cut-off, we take the $\L\ra\inf$ limit of \rf{cnl2} and expand
it in powers of $p^2/m^2$ we get
\beq
\left( \frac{1}{2m}-\frac{1}{16\pi
m}\frac{g}{m^2} \right) + \left( \frac{-1}{4m}-\frac{3}{32\pi
m}\frac{g}{m^2} \right) \frac{p^2}{m^2}+ \cdots
\eeq
we find that the resummed value \rf{c0res} is a good estimate of the
first term. We need to go to the next order to perform a resummation of
the $p^2/m^2$ coefficient in eq. \rf{cp2}, so we cannot compare it with
the $\L\ra\inf$ value.

The approach we have just described fails when $\alpha_0=0$,
which happens for example in the perturbative treatment of masslesss theories.
However the Schr\"odinger functional

\beq
\Psi_t[\f,\f']=\langle\,\f\,|\,e^{-\hat{H}t}\,|\,\f'\,\rangle
\eeq
does have a local expansion in which $1/t$ plays the role of the mass-gap, and from which the functional can be constructed for arbitrary
$\f,\,\f'$ using analyticity as before. (This was used in \cite{Marcos2}
to obtain the standard result for the one-loop beta function for Yang-Mills).
Since for large $t$
\beq
\Psi_t[\f,\f']\sim \Psi_0[\f]\,\Psi_0[\f']^*
\eeq
we could compute VEVs from the large $t_1,\,t_2$ behaviour of

\beq
\int \D\phi(\x)\,\,
\phi(\x_1)\cdot\cdot\cdot \phi(\x_n)\,\, \Psi_{t_1}[\upsilon,\phi]\,\Psi_{t_2}[\phi,\upsilon]
\eeq
using our previous method.

\section{Conclusions}

In this paper we have shown how to compute vacuum expectation values (VEVs)
from a knowledge of the vacuum wave-functional expressed as a
local expansion as would arise, for example, by solving the field theoretic
\Sc equation along the lines of \cite{Paul}-\cite{Marcos2}.
Such expansions are valid for slowly varying fields and so would
conventionally be thought of as making only a small contribution
to VEVs. However we have shown that VEVs are analytic in a momentum
cut-off, so that the large cut-off behaviour which we want to compute can be
obtained from a knowledge of the VEVs at small momenta where the
local expansion is valid.

We gave several path-integral representations of VEVs.
We
also derived a diagrammatic approach to compute VEVs from a perturbative
vacuum functional
as input, this not only showed how we recover the usual perturbative
expansion, but also how we could get the same VEVs with a dimensionally
reduced, and non-local,  Euclidean effective action ($2W[\phi]$). We have justified the local expansion of the vacuum functional in
appendix C. We have also interpreted the vacuum functional in terms of random paths, which
suggests a generalisation to String Field Theory.

We discussed in detail the non-perturbative analyticity properties
of VEVs in the cut-off, for the scalar $1+1$ theory, illustrating this
using perturbation theory. We believe that these results generalise
to higher dimensional theories and also to higher spin fields,
having shown that analyticity is unaffected by the inclusion of the boundary
counter-terms that are the new features of this generalisation.


\section*{Acknowledgments}
A. Jaramillo has been partially supported by DGES under contract
PB95-1096, by a doctoral fellowship from IVEI, by a travel grant
from the British Council and by some financial aid from the
Mathematical Sciences
Department of Durham University (where part of this work has been
performed) whose hospitality is also thanked. He wishes to acknowledge
fruitful disussions with E. Fradkin, E. Gallego, V.P. Nair, R. Perry,
R. Schiappa, A. Vainshtein, V. Vento and U.J. Wiese.
P. Mansfield is grateful to the British Council for a travel grant
and to the University of Valencia for hospitality.


\section*{Appendix A}
In this appendix we will construct an alternative representation of the
VWF which we will discuss in the context of perturbation theory. In
particular, we will derive the relation \rf{Dyson0}. Consider
\beq
|\,\Psi_0\,\rangle=\lim_{t\rightarrow\inf}e^{-(\hH-E_0)t}|\,\Psi_0^{(0)}\rangle
\eeq
where $\Psi_0^{(0)}$ is the free VWF, $\hH$ is the full Hamiltonian and
$E_0$  is the vacuum energy. This equation will show us how to compute
the VWF in perturbation theory (provided that $\langle\,\Psi_0\,|\,\Psi_0^{(0)}\,\rangle\ne
0$). We can modify this equation by introducing a new term which will
not alter the formula
\beq
|\,\Psi_0\,\rangle=\lim_{t\rightarrow\inf}e^{-(\hH-E_0)t} e^{(\hH_0-E_0^{(0)})t}
|\,\Psi_0^{(0)}\,\rangle\equiv \lim_{t\rightarrow\inf} \hat U(t)|\,\Psi_0^{(0)}\,\rangle
\label{Ueq}
\eeq
We will separate the interaction part from the free one,
$\hH=\hH_0+\hH_i$ and $E_0=E_0^{(0)}+E_0^{(i)}$ where $\hH_0$ and
$E_0^{(0)}$ are the free Hamiltonian and free vacuum energy
respectively. If we take a time derivative of $\hat U(t)$ we get
\beq
-\frac{d}{dt}\hat U(t)=\hat U(t)\, \hat V(-t)
\eeq
where $\hat V(t)\equiv e^{(\hH_0-E_0^{(0)})t}\hat V(0)
e^{-(\hH_0-E_0^{(0)})t}=e^{\hH_0t}\hat V(0) e^{-\hH_0t}$ and $\hat
V(0)\equiv \hH_i-E_0^{(i)}$. By integrating this equation between $0$
and $t$ we get
\beq
\hat U(t)={\bf 1} - \int_0^t dt'\,\, \hat U(t')\, \hat V(-t')
\label{rec}
\eeq
As usual, solving this equation by iteration gives
\beq
\hat U(t)=Te^{- \int_0^t dt'\,\, \hat V(-t')}=
Te^{- \int_{-t}^0 dt'\,\, \hat V(t')}
\eeq
Therefore
\beq
\ba{c}
{\dis |\,\Psi_0\,\rangle=Te^{- \int_{-\inf}^0 dt\,\, (\hH_i(t)-E_0^{(i)})}
|\,\Psi_0^{(0)}\,\rangle=\lim_{t\rightarrow\inf}e^{tE_0^{(i)}}Te^{ - \int_{-t}^0
dt'\,\, \hH_i(t')}|\,\Psi_0^{(0)}\,\rangle=}\\
\\
{\dis = N e^{-\ep\hH_i(0)}
e^{-\ep\hH_i(-\ep)}\cdots e^{-\ep\hH_i(-\inf)}}
\ea
\label{Dyson}
\eeq
where $\hH_i(t)$ has the same time-evolution than $\hat V(t)$. Now we
can expand the exponential to get the perturbative series.  The
conjugate relation will be
\beq
\langle\,\Psi_0\,|=\langle\,\Psi_0^{(0)}\,|Te^{- \int_0^\inf dt\,\, (\hH_i(t)-E_0^{(i)})}
=\lim_{t\rightarrow\inf}e^{tE_0^{(i)}}\langle\,\Psi_0^{(0)}\,|Te^{ - \int_0^t
dt'\,\, \hH_i(t')}
\label{conj}
\eeq
The previous relation \rf{Dyson} is easily related to the
usual Rayleigh-\Sc one if we substitute in equation \rf{rec} the
definition of $\hat U(t)$ given in \rf{Ueq}
\beq
\hat U(t)={\bf 1} - \int_0^t dt'\,\, e^{-(\hH-E_0)t'} (\hH_i-E_0^{(i)})
e^{(\hH_0-E_0^{(0)})t'}
\eeq
(notice that the only $t$-dependence is on the exponentials) which
gives
\beq
|\,\Psi_0\,\rangle=\hat U(\inf)|\,\Psi_0^{(0)}\,\rangle=|\,\Psi_0^{(0)}\,\rangle-\int_0^\inf dt\,\,
e^{-(\hH-E_0)t}(\hH_i-E_0^{(i)}) |\,\Psi_0^{(0)}\,\rangle
\eeq
We realize that the exponential becomes $1$ when it is applied to
$|\,\Psi_0\,\rangle$ and therefore we will substitute $\hH-E_0$ by $\hH-E_0+i\ep$
and we will demand that the limit $\ep\rightarrow 0$ will give a finite
value. Now, we perform the $t$-integration
\beq
|\,\Psi_0\,\rangle=|\,\Psi_0^{(0)}\,\rangle-\frac{1}{\hH-E_0+i\ep} (\hH_i-E_0^{(i)})
|\,\Psi_0^{(0)}\,\rangle
\label{vacpert}
\eeq
$E_0^{(i)}$ will be given by the condition that the limit
$\ep\rightarrow 0$ should be finite (in other words,
$\langle\,\Psi_0\,|\hH_i-E_0^{(i)}|\,\Psi_0^{(0)}\rangle=0$). This is essentially the
Lippmann-Schwinger equation (which here also applies to the discrete
part of the spectrum).
Now, if we insert $\sum_n
|\,\Psi_n^{(0)}\,\rangle\langle\,\Psi_n^{(0)}\,|$ after the $\frac{1}{\hH-E_0+i\ep}$ term
and we perform an expansion in the coupling constant $\l$ ($\hH_i\equiv
\l \hH_{int}$) we get the usual Rayleigh-\Sc perturbation expansion
\cite{Hatfield}. For example, to first order we will obtain
\beq
\ba{l}
{\dis \Psi_0[\phi]=\Psi_0^{(0)}[\phi]+\l \sum_{n\ne 0}\int d\kk_1\cdots
d\kk_n \,\,}\\
\\
{\dis \frac{\langle\,\Psi_n^{(0)}(\kk_1,\cdots,\kk_n)\,|\hH_{int}|\,\Psi_0^{(0)}\rangle}
{E_0^{(0)}-E_n^{(0)}}\Psi_n^{(0)}[\phi;\kk_1,\cdots,\kk_n]}
\ea
\label{pert}
\eeq
where we have taken the $\ep\rightarrow 0$ limit. This relation hides
the diagrammatic method of section 3, but at the end of that section
this equivalence is shown using the
previous formula \rf{Dyson}.

Of course, in order to derive the perturbative series \rf{pert} we
could directly expand
\beq
\Psi_0[\phi]=\lim_{t\rightarrow\inf}e^{tE_0^{(i)}}\langle\,\phi(\x)\,|Te^{ -
\int_{-t}^0 dt'\,\, \hH_i(t')}|\,\Psi_0^{(0)}\,\rangle
\label{Dyson2}
\eeq
getting to first order
\beq
\ba{c}
{\dis
\Psi_0^{(0)}[\phi]-\l \int_{-t}^0 dt'\,\,
\langle\,\phi(\x)\,|e^{t'\hH_0}\hH_{int}e^{-t'\hH_0}|\,\Psi_0^{(0)}\,\rangle=
\Psi_0^{(0)}[\phi]-}\\
\\
{\dis -\l \int_{-t}^0 dt'\,\,
\sum_{n,m=0}^\inf
e^{t'(E_n^{(0)}-E_m^{(0)})}\langle\,\Psi_n^{(0)}\,|\hH_{int}|\,\Psi_m^{(0)}\,
\rangle
\langle\,\Psi_m^{(0)}\,|\,\Psi_0^{(0)}\,\rangle \Psi_n^{(0)}[\phi] }
\ea
\eeq
When both, the $t$-integral and the limit, are done we get
\beq
\ba{c}
{\dis \Psi_0^{(0)}[\phi]-\l t\langle\,\Psi_0{(0)}\,|\hH_{int}|\,\Psi_0^{(0)}\,\rangle
\Psi_0^{(0)}[\phi] -}\\
\\
{\dis -\l \sum_{n\ne 0} \frac{1}{E_n^{(0)}-E_0^{(0)}}
\langle\,\Psi_n^{(0)}\,|\hH_{int}|\,\Psi_0^{(0)}\,\rangle \Psi_n^{(0)}[\phi] }
\ea
\eeq
Finally
\beq
\ba{c}
{\dis\Psi_0[\phi]= \lim_{t\rightarrow\inf}e^{tE_0^{(i)}} \bigg\{
(1-\l t \langle\,\Psi_0^{(0)}\,|\hH_{int}|\,\Psi_0^{(0)}\,\rangle)\Psi_0^{(0)}[\phi] + }\\
\\
{\dis +\l \sum_{n\ne 0} \frac{1}{E_0^{(0)}-E_n^{(0)}}
\langle\,\Psi_n^{(0)}\,|\hH_{int}|\,\Psi_0^{(0)}\,\rangle \Psi_n^{(0)}[\phi] \bigg\} }
\ea
\eeq
As before, we choose $E_0^{(i)}$ in such a way that the limit exists
(although that this time is another limit) and
therefore if we take $E_0^{(i)}=\l
\langle\,\Psi_0^{(0)}\,|\hH_{int}|\,\Psi_0^{(0)}\,\rangle$ the $t$-limit will be finite
until $O(\l^2)$. As we can see, we got the same result than in
equation \rf{pert}.


\section*{Appendix B}
In this appendix we will discuss the mathematical details of the resummation
program through the use of several examples. Let
us begin with the integral
\beq
K(\L)=\int_0^\L dx\,\,\frac{1}{1+x^2}=\arctan(\L)
\label{arctg}
\eeq
$K(\L)$ is analytic in the cut $\L$-plane (with the cut chosen from
$\pm i$ to $\inf$). We want to compute $K(\inf)$ from a series expansion for
small $\L$. For $\L<1$ we get
$$K(\L)=\sum_{n=0}^\inf \frac{(-1)^n}{2n+1}\L^{2n+1}$$
We can compute its Borel transform by
constructing a new series with a $(2n+1)!$ in the denominator.
\beq
I(\l)=\sum_{n=0}^\inf \frac{(-1)^n}{2n+1}\frac{\l^{2n+1}}{(2n+1)!}=
\int_0^\l dx\,\,\frac{\sin(x)}{x}
\label{Barctg}
\eeq
This is the analogous step to that from eq. \rf{s-1} to \rf{Ilam} but with a different
variable. Although an
integral representation similar to \rf{resum} is known (see \cite{Borel}
for details on the Borel summation of series), it is not used. Instead,
$K(\L)$ is usually recovered from $I(\l)$ through the ``inverse Borel
transform''
\beq
K(\L)=\int_0^\inf d\l\,\, I(\L\l)\, e^{-\l}
=\frac{1}{\L}\int_0^\inf d\l\,\, I(\l)\, e^{-\l/\L}
\label{Binv}
\eeq
If $I(\l)$ is given in terms of a series (as in our discussion of VEVs), then the
integral whould trivially remove the $n!$ term and we would get back
the previous series. But usually with Borel-summable asymptotic series one
only knows $I(\l)$ in a series with
finite convergence radius which is insufficent to compute the infinite integral
\rf{Binv}. One then has to find $I(\l)$ for larger
$\l$ by analytic continuation and this is in general does not yield a polynomial.
Conformal mapping and Pad\'e approximants are used for that purpose. In our
case, we work with series $K(\L)$ with finite convergence radius which
implies that $I(\l)$ series will be valid for all the integration domain, so
the analytic continuation cannot provide us with a non-polynomial form and
we have to change the method to recover $K(\L)$. We solve the problem by
using the integral representation \rf{resum} which allows us to compute
$K(\L)$ from $I(\inf)$ as we will show in an example below. We prefer this
method over the conformal mapping or the Pad\'e approximants because it is
more systematic to implement in a field theoretical large distance
expansion, where a series in the cut-off naturally arises.

If $I(\l)$ has a finite $\L\ra\inf$ limit, eq. \rf{Binv} gives
$K(\inf)=I(\inf)$, which we apply to our example \rf{Barctg}
\beq
K(\inf)=\int_0^\inf dx\,\,\frac{\sin(x)}{x}=\frac{\pi}{2}
\eeq
agreeing with the $K(\inf)$ computed directely from \rf{arctg}. Usually
we will have to compute $I(\inf)$ approximately, by truncating the series.
The truncation error in this alternating series is bounded by the first
neglected term. But we have treated the exact expression to be able to study the convergence.

We can adapt our method to the case
of $K(\L)$ for finite $\L$, greater than $1$. Consider
\beq
K(\L,s)=\sqrt{s}\int_0^{\L/\sqrt{s}}dx\,\,\frac{1}{1+x^2}
\label{Ks1}
\eeq
which, as a function of $s$, is analytic in the cut $s$-plane (with the cut
from $-\L^2$ to $0$). Then we expand $K(\L,s)$ into powers of $(s-1)^{-n}$
(we have analyticity for large $s$) as in eq. \rf{s-1}.
\beq
K(\L,s)=\L+\sum_{n=1}^\inf (-1)^n \frac{P_{2n+1}(\L)}{(s-1)^n}
\eeq
with $P_{2n+1}(\L)=\int_0^\L dx\,\, x^2 (1+x^2)^{n-1}$.
Although the point $s=1$ (where we recover $K(\L)$) is beyond the
convergence radius, we will use the relation \rf{Ilam}, written as
\beq
I(\l)=\L + \sum_{n=1}^\inf (-1)^n P_{2n+1}(\L)\frac{\l^n}{n!}.
\eeq
Again, we can obtain the whole sum in a closed form
\beq
I(\l)=\L + \int_0^\l dx\,\, \frac{e^{-x}}{4x^{3/2}}(2\sqrt{x}
e^{-x\L^2}\L-\sqrt{\pi}\textrm{erf}(\sqrt{x}\L))
\eeq
where $\textrm{erf}(x)$ is the error function. The limit $\l\ra\inf$
can also be obtained exactly
\beq
\lim_{\l\ra\inf}I(\l)=\arctan(\L)
\eeq
using this result into the relation \rf{recov} gives
$K(\L)=I(\inf)=\arctan(\L)$, which agrees with \rf{arctg}.

As another example we will compute the integral
\beq
K(\L)=\int\limits_{|k|<\L} \frac{dk}{2\pi} \left( \frac{2}{1+k^2} \right)^2
\label{gsint}
\eeq
for $\L\ra\inf$, which corresponds to the normalization of the
wave-function \rf{Dgs}. Its Borel transform is given by
\beq
I(\l)=\frac{4}{\pi}\sum_{n=0}^\inf (-1)^n \frac{n+1}{2n+1}\frac{\l^{2n+1}}
{(2n+1)!}=\frac{2}{\pi}(\sin(\l)+\textrm{Si}(\l))
\eeq
Where $\textrm{Si}$ denotes the sine integral function. We realize that
there is no $\l\ra\inf$ limit of $I(\l)$ because of the oscillatory behavior.
The same thing happens with the $\langle\,\Psi_0\,|k^2|\,\Psi_0\,\rangle$ integral
\beq
K(\L)=\int\limits_{|k|<\L} \frac{dk}{2\pi} \left( k^2 \frac{2}{1+k^2} \right)^2
\label{gsk2int}
\eeq
which has the Borel transform
\beq
I(\l)=\frac{2}{\pi}(\textrm{Si}(\l)-\sin(\l))
\eeq
If we take a truncated series for $K(\L)$  then we
will estimate $I(\inf)$ as the value of the approximant at its first
stationary point (if we take the highest $\l^n$ term to have $n$ even then
the  stationary point will be a local maximum). For the case of
$|\tilde\Psi_0|^2$ to $O(k^2)$ we get that \rf{gsint} gives $1.47$ (to be
compared with 1, the $\L\ra\inf$ value). For \rf{gsk2int} with the
wave-function squared expanded to $O(k^2)$ we get $0.89$ (to be also
compared with $\langle\,\Psi_0\,|k^2|\,\Psi_0\,\rangle=1$). Higher order terms
will give worse estimates due to the fact that they are estimating the first
maximum (the $O(k^6)$ truncated $|\tilde\Psi_0|^2$ gives $1.56$, the
$O(k^8)$ gives $1.7$ like all higher approximants).

We can ask when does this oscillation occur?
From eq. \rf{Binv} we can see that
if we substitute $I(\l)$, for large $\l$, by $e^{i a\l}$ then we find that
$K(\L)$ has to have a pole at $\L=-i/a$. To solve this no-convergence
problem we will do a mapping, to move the pole. In fact, in eq. \rf{Ks1} we
may think that we have performed a mapping $\L\ra \mu/(1+1/z)^{1/2}$ with
$\mu=\L$, but now the pole has moved to the negative real axis. But we
should keep $\L$ finite. Then the idea to compute \rf{gsint} would be to
choose first a high value of $\L$ (for instance, $\L=4$ where $K(4)=0.994$),
then we use eq. \rf{resum} to eq. \rf{recov} to compute $K(4)$ (as we did
for \rf{Ks1}). We get that
$K(\L,s)=\sum A_n(\L)s^{-n}=\sum B_n(\L)(s-1)^{-n}$ and then
$I(\l)=\sum B_n(\L)\l^n/n!$. But because $B_n(4)\sim 18^n (-1)^n$, we will
have to truncate the series at least at order $n=46$. Therefore we have to
improve the method. In fact an easy solution is the one used in section 5. Consider
\beq
K(s)=\int\limits_{|k|<1/\sqrt{s}} \frac{dk}{2\pi}
\left( \frac{2}{1+k^2} \right)^2
\label{Ksqrt}
\eeq
to compute the $s\ra 0$ limit we will perform its Borel transform,
which we define
\beq
I(\l)=\frac{1}{2\pi i}\int_C \frac{ds}{s}\, K(s) \, e^{\l s}
\eeq
where C is a large, almost closed, circular (counterclockwise) path which
does not cross the negative real axis. With the help of
\beq
\frac{1}{\G(z)}=\frac{1}{2\pi i}\int_C dx\,\, x^{-z}\, e^{x}
\eeq
(where the countour C surrounds the negative real axis clockwise), we see
that the Borel transform of
\beq
K(s)=\sum_{n=0}^\inf (-1)^n a_n \frac{1}{s^{n+1/2}}
\eeq
is given by
\beq
I(\l)=\sum_{n=0}^\inf (-1)^n a_n \frac{\l^{n+1/2}}{\G(n+3/2)}
\label{qmn}
\eeq
We see that we have divided by $(n+1/2)!\equiv \G(n+3/2)$. This series has a
good convergence (by only taking two terms we get $K(0)=I(\inf)=0.83$ as
estimate), but the function $I(\l)$ has one oscillation before decaying. It
has a maximum at $\l=2.24$ and $I(2.24)=1.042$. If we estimate $I(\inf)$ by
its value at the first local maximum, then $1.042$ will be our result when
computing the series truncated at any $n>7$. Therefore we have to truncate
\rf{qmn} for $n=3$ getting $I(\inf)$=0.978 which is close enough to $K(0)=1$.

We have used our resumation method to compute finite integrals. If we
consider a divergent integral, then the integral will remain divergent
after resummation as we can see with the following example
\beq
K(\L)=\int_0^\L dx\,\, \frac{1}{1+x}
\eeq
we can compute $I(\L)$ in a closed form
\beq
I(\l)=\int_0^\l dx\,\, \frac{1-e^{-x}}{x}
\eeq
and we see that $I(\L)\sim \log(\L)$ for large $\L$, reproducing the
previous divergence.


\section*{Appendix C}
In this appendix we will give argument for the validity of the local
expansion of the VWF for a scalar theory with a mass gap. Again, it
would be very
convenient to use a path integral representation for the VWF.
\beq
|\Psi_0[\phi]|^2=\int\D J(x)\,\,\exp(-G_c[J(x)\d(t)]-i\int dx\,\,
J(x)\phi(x))
\label{frad}
\eeq
where $G_c[J(x,t)]$ is the generator of connected Green's functions in
Euclidean space. This formula can be derived \cite{Fradkin} by interpreting
$e{-G_c[J(x)\d(t)]}$ (in terms of $\Psi_0[\phi]$) as the
(functional) Fourier transform of $|\Psi_0[\phi]|^2$. When we Fourier
transform back we get \rf{frad}. The relation \rf{frad} has the advantage
that we can think of $|\Psi_0[\phi]|^2$ as a partition function with a
non-local action $G_c[J(x)\d(t)]$. Because we expect that the connected
Green's functions will be analytic at zero momenta (with the nearest pole
given by the mass), they will have an exponential decay in configuration
space (even if we put $t=0$) which implies that the equal-time connected
Green's functions in momentum space will be analytic in a neighborhood of
zero. This applies even if we had a massless Lagrangian (but non-zero mass
gap). Because the logarithm of the square of the VWF
is the generator of the
connected Green's functions (of the field $J$), if we have slowly varying
$\phi$-configurations (low external momenta) we can substitute the non-local
action by a local action without changing the infrared amplitudes (i.e.
giving the same slowly varying $|\Psi_0[\phi]|^2$). Thanks to this assumed
universality, we can now work with an effective field theory (for the quantum
field $J(x)$) in the usual way. We see that  the connected
two-point function at zero momentum will give a mass term for the field $J$.
Thanks to this mass term, we will have analyticity at low momenta of the
$J$-Green's functions which means that we can perform the local expansion of
the VWF. The fact that the effective theory is non-renormalizable does not
cause trouble as it is common with effective field theories. But we still
have to assume that the quantum corrections to the mass are small. Finally,
we should
realize that although the VWF is finite when we
remove the cut-off (and then also
the coefficients of its local expansion), the coefficients of the effective
action are (generally) not finite (because they take into account the change
in the theory at short distances due to the cut-off).




\end{document}